%% file: sdhvp.tex
\documentclass[12pt,nofootinbib,tightenlines,preprintnumbers]{revtex4}
\usepackage{graphicx}

\usepackage{amssymb}
\usepackage{amsmath}
\usepackage{xcolor}
\usepackage{booktabs}
\usepackage{placeins}
\usepackage{hyperref}
\usepackage[utf8]{inputenc}
\hypersetup{%
        pdftitle    = {Hadronic vacuum polarization in the muon g-2: The short-distance contribution from lattice QCD},
}%

\input{definitions}

\def\af#1#2{a_\mu^{{#1},{#2}}}
\def\Pif#1#2{\Pi^{({#1},{#2})}}
\def\Gf#1#2{G^{({#1},{#2})}}

\newcommand{\aID}{(a_\mu^{\rm hvp})^{\rm ID}}
\newcommand{\aSD}{(a_\mu^{\rm hvp})^{\rm SD}}
\newcommand{\aSDO}{(a_\mu)^{\rm SD}}
\def\aSDf#1#2{(a_\mu^{{#1},{#2}})^{\rm SD}}
\newcommand{\aSDsub}{(a_\mu^{\rm hvp})^{\rm SD}_{\rm sub}}
\def\aSDsubf#1#2{(a_\mu^{{#1},{#2}})^{\rm SD}_{{\rm sub}}}
\newcommand{\Qref}{Q_{\rm ref}}
\newcommand{\bSD}{b}
\def\bSDf#1#2{\bSD^{({#1},{#2})}}

\begin{document}
\begin{titlepage}
\begin{flushright}
MITP-24-011\\
CERN-TH-2024-011
\end{flushright}

\vskip 0.5 cm
\begin{center}
  {\Large\bf Hadronic vacuum polarization in the muon $g-2$:\\ The short-distance contribution from lattice QCD \\[0.5ex]}
\end{center}
\vskip 1.0cm
\begin{center}
{\large
  Simon Kuberski$^{a}$,
  Marco C\`e$^{b}$,
  Georg von Hippel$^{c}$,
  Harvey~B.~Meyer$^{c,\,d,\,e}$,
  Konstantin Ottnad$^{c}$,
  Andreas Risch$^{f,\,g}$ and
  Hartmut Wittig$^{c,\,d,\,e}$
}
\vskip 0.5cm
$^{a}$\,Department of Theoretical Physics, CERN, 1211 Geneva 23, Switzerland
\vskip 0.15cm
$^{b}$\,Dipartimento di Fisica, Universit\`a di Milano-Bicocca and INFN, Sezione di Milano-Bicocca, Piazza della Scienza 3, 20126 Milano, Italy 
\vskip 0.15cm
$^{c}$\,PRISMA$^+$ Cluster of Excellence and Institut f\"ur Kernphysik, Johannes Gutenberg-Universit\"at Mainz, 55099 Mainz, Germany
\vskip 0.15cm
$^{d}$\,Helmholtz-Institut Mainz, Johannes Gutenberg-Universit\"at Mainz, 55099 Mainz, Germany
\vskip 0.15cm
$^{e}$\,GSI Helmholtz Centre for Heavy Ion Research, 64291 Darmstadt, Germany
\vskip 0.15cm
$^{f}$\,Department of Physics, University of Wuppertal, Gaussstr. 20, 42119 Wuppertal, Germany
\vskip 0.15cm
$^{g}$\,John von Neumann-Institut f{\"u}r Computing NIC, Deutsches Elektronen-Synchrotron DESY, Platanenallee 6, 15738 Zeuthen, Germany
\end{center}

\vskip 0.7cm

\noindent
We present results for the short-distance window observable of the
hadronic vacuum polarization contribution to the muon $g-2$, computed
via the time-momentum representation (TMR) in lattice QCD. A key
novelty of our calculation is the reduction of discretization effects
by a suitable subtraction applied to the TMR kernel function, which
cancels the leading $x_0^4$-behaviour at short distances. To
compensate for the subtraction, one must substitute a term that can be
reliably computed in perturbative QCD. We apply this strategy to our
data for the vector current collected on ensembles generated with $2+1$
flavours of O($a$)-improved Wilson quarks at six values of the lattice
spacing and pion masses in the range $130-420$\,MeV. Our estimate at
the physical point contains a full error budget and reads
$\aSD=68.85(14)_{\rm stat}\,(42)_{\rm syst}\cdot10^{-10}$, which
corresponds to a relative precision of 0.7\%. We discuss the
implications of our result for the observed tensions between lattice
and data-driven evaluations of the hadronic vacuum polarization.

\vfill
\centerline{January 22, 2024}

\end{titlepage}

 \newpage
 \tableofcontents
 \newpage

\input{sec_intro}
\input{sec_definitions}
\input{sec_lattice_setup}

\input{sec_results}

\input{sec_conclusions}

\acknowledgments{We thank Rainer Sommer for discussions on the cutoff
  effects of correlators at short distances and Kohtaroh Miura for his
  contribution to the computation of finite-size effects.  Calculations 
  for this project have been performed on the HPC clusters Clover and
  HIMster-II at Helmholtz Institute Mainz and Mogon-II at Johannes
  Gutenberg-Universität (JGU) Mainz, on the HPC systems JUQUEEN and
  JUWELS
  and on the GCS Supercomputers HAZELHEN and HAWK at
  Höchstleistungsrechenzentrum Stuttgart (HLRS). The authors
  gratefully acknowledge the support of the Gauss Centre for
  Supercomputing (GCS) and the John von Neumann-Institut für Computing
  (NIC) for projects HMZ21, HMZ23 at JSC and project GCS-HQCD at
  HLRS. This work has been supported by Deutsche
  Forschungsgemeinschaft (German Research Foundation, DFG) through
  Project HI~2048/1-2 (Project No.\ 399400745) and through the Cluster
  of Excellence ``Precision Physics, Fundamental Interactions and
  Structure of Matter'' (PRISMA+ EXC 2118/1), funded within the German
  Excellence strategy (Project No.\ 39083149). This project has
  received funding from the European Union’s Horizon Europe research
  and innovation programme under the Marie Sk\l{}odowska-Curie grant
  agreement No 101106243. The research of M.C. is funded through the
  MUR program for young researchers ``Rita Levi Montalcini''.  We are
  grateful to our colleagues in the CLS initiative for sharing
  ensembles.}

\clearpage
\appendix
\input{app_tables}
\FloatBarrier
\bibliographystyle{jhep_collab}
\bibliography{biblio}

\end{document}

%% file: definitions.tex
\newcommand{\be}{\begin{equation}}
\newcommand{\ee}{\end{equation}}
\newcommand{\ba}{\begin{eqnarray}}
\newcommand{\ea}{\end{eqnarray}}
\newcommand{\la}{\label}

\newcommand{\<}{\langle}
\renewcommand{\>}{\rangle}

\newcommand{\txts}{\textstyle}

\newcommand{\MeV}{{\rm{MeV}}}

\newcommand{\cv}{c_{\rm V}}

\newcommand{\ahvp}{a_\mu^{\rm hvp}}

\newcommand{\bea}{\begin{eqnarray}}
\newcommand{\eea}{\end{eqnarray}}

\newcommand{\psib}{\overline{\psi}}

\newcommand{\amu}{a_\mu^{\mathrm{hvp}}}

\newcommand{\loc}{{\scriptstyle\rm L}}
\newcommand{\cons}{{\scriptstyle\rm C}}
\newcommand{\mloc}{ {\scriptscriptstyle\rm L} }
\newcommand{\mcons}{ {\scriptscriptstyle\rm C} }

\newcommand{\Xpi}{\Phi_2}
\newcommand{\XK}{\Phi_4}
\newcommand{\Xdelta}{\Phi_{\delta}}
\newcommand{\Xa}{X_a}

%% file: sec_intro.tex
\section{Introduction}

The anomalous magnetic moment of the muon, as a low-energy precision observable, has long served as a test of the
Standard Model (SM) of particle physics.
Its  experimental measurement has reached a spectacular precision. With the latest result of
the Fermilab $(g-2)$ experiment at the 0.20\,ppm level~\cite{Muong-2:2023cdq}, the stakes
for the SM-based theoretical prediction have been raised even further.
The 2020 White Paper (WP2020) of the Muon $(g-2)$ Theory Initiative~\cite{Aoyama:2020ynm} arrived at a result with a quoted precision
of 0.37\,ppm, the uncertainty being dominated by the hadronic vacuum polarization (HVP) contribution and to a lesser extent
by the hadronic light-by-light contribution. The very recent experimental result~\cite{Muong-2:2023cdq}
is in tension with the 2020 prediction at the level of five standard deviations.
However, several results and observations of the last three years concerning the HVP contribution
have cast doubt on the reliability of the 2020 prediction~\cite{Aoyama:2020ynm}.

First, while the  WP2020 evaluation of the HVP contribution $\ahvp$ was based solely on the
dispersion relation relating it to the inclusive $e^+e^-\to {\rm hadrons}$ cross section,
a lattice QCD calculation with subpercent precision published by the BMW collaboration in 2021~\cite{Borsanyi:2020mff}
arrived at a result larger by 2.1 combined standard deviations than the WP2020 one.
If one were to replace the WP2020 evaluation of the HVP contribution by that of the BMW collaboration,
the tension between the SM and the latest experimental result for the muon $(g-2)$ would shrink to $1.7\,\sigma$.

Secondly, four lattice collaborations~\cite{Borsanyi:2020mff,Ce:2022kxy,ExtendedTwistedMass:2022jpw,RBC:2023pvn}
have independently computed a partial HVP contribution known as the `window' or `intermediate-distance' contribution $\aID$,
consistently obtaining a  result larger by about three percent than via the dispersive method:
the tensions between the individual lattice calculations
and the dispersion evaluation~\cite{Colangelo:2022vok} range between $3.7\,\sigma$ and $3.9\,\sigma$.
Even more results~\cite{Borsanyi:2020mff,Lehner:2020crt,Wang:2022lkq,Aubin:2022hgm,Ce:2022kxy,ExtendedTwistedMass:2022jpw,FermilabLatticeHPQCD:2023jof,RBC:2023pvn}
are available for partial flavour contributions, which further corroborate the discrepancy~\cite{Benton:2023dci}.

Thirdly, a new measurement of the $e^+e^-\to \pi^+\pi^-$ cross section by the CMD-3 collaboration has appeared, which lies
higher than the previous measurements. This new result is very significant,
since in the dispersive evaluation of $\ahvp$, the $\pi^+\pi^-$ channel dominates,
with the $\rho$ meson region $600\,{\rm MeV}<\sqrt{s}<900\,{\rm MeV}$ providing more than half of $\ahvp$.
If confirmed, the CMD-3 result would bridge the difference between the WP2020 and the BMW calculations of $\ahvp$.

The `intermediate window' alluded to above is the second of three
subcontributions in a partition of $\ahvp$ based on the Euclidean time separation between two electromagnetic
currents~\cite{Bernecker:2011gh,Blum:2018mom}.
The first of these subcontributions is the short-distance quantity $\aSD$ extending up to 0.4\,fm (with a soft edge), which is the focus
of this paper, while the third, numerically dominant part is the long-distance contribution $(a_\mu^{\rm hvp})^{\rm LD}$.
The short-distance quantity only represents about ten percent of $\ahvp$; in this respect,
its determination at a precision level of two percent would currently be sufficient. However, 
we argue below that it is worth aiming at a subpercent precision for $\aSD$, in order to 
provide a clue as to the origin of the well established discrepancy in the intermediate window quantity.

If one assumes that the tension between the lattice and the dispersive
determinations of the intermediate window is due to an underestimate
(by an overall factor of about 0.94) of the experimentally determined $R(s)$ ratio in the
interval from 600 to 900\;MeV, then one expects to find an
underestimate of the dispersively determined short-distance quantity
by more than one percent.  Specifically, based on our own previous
calculation~\cite{Ce:2022kxy} and the estimates given therein of the
fractional contributions to $\aSD$ and $\aID$ associated with the
interval $600\,{\rm MeV}<\sqrt{s}<900\,{\rm MeV}$, the expectation
would be an excess of $(1.44\pm0.37)\%$ of our lattice result for
$\aSD$ over the dispersive one. Thus it is clear that a  precision well under one percent
is required on $\aSD$ in order for such a calculation to have an impact.
The existing lattice results~\cite{ExtendedTwistedMass:2022jpw,RBC:2023pvn} for $\aSD$ are consistent
with the scenario above; at the same time, the evidence for a discrepancy with the dispersive value for $\aSD$
lies only at the $1.4\,\sigma$ level even for the most precise result~\cite{ExtendedTwistedMass:2022jpw}.

There are further hints in favour of the scenario of an underestimated $R(s)$ ratio in the $\rho$ region.
One is the calculation of the hadronic contribution to the running of the electromagnetic coupling $\alpha(q^2)$.
The Mainz/CLS publication of 2022~\cite{Ce:2022eix} showed that a tension develops between photon
virtualities of zero and 1\,GeV$^2$, while the running 
is consistent with the dispersive approach at higher $q^2$; see~\cite{Borsanyi:2017zdw,Borsanyi:2020mff}
for the earlier calculation by the BMW collaboration. This observation points to an origin
at  $\sqrt{s}\lesssim1\,{\rm GeV}$ in the dispersive approach, or alternatively to an issue with
the lattice data at Euclidean times $x_0\gtrsim1\,{\rm fm}$.
Secondly, the fact that the absolute difference in $\aID$ appears to reside entirely in the light-quark
connected contribution~\cite{Benton:2023dci} can be interpreted as a further consistency check, since
the on-shell effects of the strange quark start at $\sqrt{s}=2m_K\simeq 990\,{\rm MeV}$.
Thirdly, a dedicated study of the  spectral function, smeared by a Gaussian kernel,
has recently been published~\cite{ExtendedTwistedMassCollaborationETMC:2022sta},
where the strongest tension (about $2.9\,\sigma$) between the lattice and the $e^+e^-$ data-based
result is seen around the $\rho$ peak with a broad smearing kernel of width 630\,MeV.
Last but not least, an increase of the phenomenological $R(s)$ ratio by six percent in the interval
$600\,{\rm MeV}<\sqrt{s}<900\,{\rm MeV}$ would bring the WP2020 prediction for the muon $(g-2)$
into perfect agreement with the direct experimental measurement. For all these reasons,
we consider such a scenario to be a useful working hypothesis. At the same time,
in this scenario a very convincing explanation (within or outside the Standard Model)
would have to be found why all previous experimental
measurements of the $\pi^+\pi^-$ channel and possibly also $\pi^+\pi^-\pi^0$ channel
are systematically suppressed in the $\rho,\omega$ region, well beyond what the quoted uncertainties
would allow for. 

The short-distance observable $\aSD$ can be obtained very precisely in
lattice QCD as far as statistical errors are concerned. Some sources
of uncertainty that are very relevant to the calculation of $\ahvp$,
such as finite-volume effects or a slight mistuning of the light
quark masses, play hardly any role in $\aSD$. Also the sensitivity to
the `absolute scale setting', meaning the calibration of the lattice
spacing in physical units, is very subdominant.  The main source of
systematic uncertainty is the presence of enhanced cutoff effects,
which must be removed via a careful continuum extrapolation if one
targets subpercent precision. Given that cutoff effects are enhanced
at short distances, we devise a specific observable that we subtract
from $\aSD$ to facilitate the continuum limit, and evaluate this
observable with the help of massless perturbation theory to order $\alpha_s^4$.
It is designed to only receive contributions from virtualities of order $Q^2$,
where $Q^2$ is an adjustable parameter chosen to lie in the perturbative regime.

This paper is organized as follows.
In section~\ref{sec:prelims}, we describe our notation and computational strategy,
our gauge ensembles and fitting procedures.
Section~\ref{s:results} contains our results for the various contributions computed using lattice QCD,
with two further, small effects evaluated in its last subsection~\ref{s:further_contribs}.
We present our final result in section~\ref{s:concl}, and discuss its impact on the current $(g-2)$ puzzle.

%% file: sec_definitions.tex
\section{Preliminaries}\label{sec:prelims}

Here we recall some basic definitions and describe our overall
computational strategy. In particular, we introduce and discuss the
decomposition of the short-distance window observable, which allows us
to gain good control over the extrapolation to the continuum limit
through a combination of perturbation theory and lattice calculations.

The electromagnetic current correlator in the time-momentum representation (TMR) is defined by
\be
\delta_{kl}\, G(x_0) = - \int d^3x\; \<J_k(x) J_l(0)\>,
\ee
where $J_\mu(x) = \frac{2}{3} \bar u \gamma_\mu u - \frac{1}{3} \bar d \gamma_\mu d - \dots$
In this representation, the hadronic vacuum polarization contribution to the muon anomalous magnetic
moment is given by\footnote{The kernel $\tilde K(x_0)$ used in our previous publication~\cite{Ce:2022kxy}
corresponds to $\tilde K(x_0) = {\rm K}(m_\mu x_0) / m_\mu^2$.}
\be
\ahvp= \Big(\frac{\alpha}{\pi m_\mu}\Big)^2 \int_0^\infty dx_0\; {\rm K}(m_\mu x_0)\; G(x_0).
\ee
The function ${\rm K}(z)$ is given analytically in Ref.\ \cite{DellaMorte:2017dyu}.
Its asymptotic properties are
\be \la{eq:Kasympt}
   {\rm K}(z) \sim \left\{\begin{array}{l@{\quad}l}
     \frac{\pi^2}{9} z^4 & z\ll 1,
     \\
     2\pi^2 z^2 & z\gg 1. \phantom{\Big|}
     \end{array}\right.
\ee
The short-distance window contribution is given by
\be\la{eq:SDdef}
\aSD 
= \Big(\frac{\alpha}{\pi m_\mu}\Big)^2 \int_0^\infty dx_0\; {\rm K}(m_\mu x_0)\; w_{\rm SD}(x_0)
\,G(x_0),
\ee
where
\ba
w_{\rm SD}(x_0) &=& 1-\Theta(x_0,d,\Delta),
\\
\Theta(x_0,d,\Delta) &\equiv& \frac{1}{2}\Big( 1 + \tanh \frac{x_0-d}{\Delta}\Big)
\ea
is a smooth step function interpolating around $x_0\approx d$ between zero and unity on a distance scale $\Delta$,
and we choose the currently standard values
\be\la{eq:stdwindparams}
d = 0.4\,{\rm fm}, \qquad \Delta = 0.15\,{\rm fm}.
\ee

For the perturbative evaluation of high-momentum contributions to $\amu$, it is useful to introduce 
the Adler function, defined as the logarithmic derivative of the vacuum polarization function,
\be
D(Q^2) = 12\pi^2\, Q^2\, \frac{d\Pi}{dQ^2}\;.
\ee
Note that $D(Q^2)$ is normalized so as to have the same high-energy asymptotics as the $R$
ratio, $D(Q^2)\rightarrow N_c \sum_f {\cal Q}_f^2$.
The quantity
\be\la{eq:Del2}
\Pi(Q^2) - \Pi(Q^2/4) = \frac{1}{12\pi^2}\int_{Q^2/4}^{Q^2} \frac{d\tau}{\tau}\; D(\tau)
\ee
will play an important role in the following.
It can be computed in the Euclidean theory via the integral~\cite{Ce:2021xgd}
\be
\Pi(Q^2) - \Pi(Q^2/4) = \frac{16}{Q^2} \int_0^\infty dx_0\;G(x_0)  \sin^4\Big(\frac{Q x_0}{4}\Big).
\ee
Note that the TMR kernel to compute this quantity is positive-definite, proportional to $x_0^4$ at short distances
and remains bounded at long distances.

\subsection{Flavour structure of the current-current correlator}
We write the current with the help of a  matrix $T^a$ acting in flavour space,
\be
J_\mu^a = \bar\psi T^a \gamma_\mu \psi, \qquad
\bar\psi = \left( \bar u ~\; \bar d ~\;\bar s ~\;\bar c ~\;\bar b  \right),
\ee
and introduce the flavour-specific TMR correlator
\be
\delta_{kl}\, \Gf{a}{b}(x_0) = - \int d^3x\; \<J_k^a(x) J_l^b(0) \>.
\ee
For $a=1,\dots 8$, we define $T^a = {\txts\frac{\lambda^a}{2}}\oplus 0_{\rm c} \oplus 0_{\rm b}$
to be given by the corresponding Gell-Mann matrix $\lambda^a$
acting in the $(u,d,s)$ sector. We also use $J_\mu^{\rm c} = \bar c \gamma_\mu c $ ($J_\mu^{\rm b} = \bar b \gamma_\mu b$) to denote the charm (bottom) current, i.e.\  $T^{\rm c} = {\rm diag}(0,0,0,1,0)$ and $T^{\rm b} = {\rm diag}(0,0,0,0,1)$. The index $\gamma$ is used
to indicate the physical quark charge matrix, $T^\gamma = {\rm diag}(\frac{2}{3},-\frac{1}{3},-\frac{1}{3},\frac{2}{3},-\frac{1}{3})$.

With this flexible notation in place, we can write the electromagnetic-current correlator as 
\be
\Gf{\gamma}{\gamma} = \Gf{3}{3} + {\txts\frac{1}{3}} \Gf{8}{8}  + {\txts\frac{4}{9}}\Gf{\rm c}{\rm c}_{\rm conn}
+   {\txts\frac{2}{3\sqrt{3}}}\Gf{\rm c}{8} + {\txts\frac{4}{9}} \Gf{\rm c}{\rm c}_{\rm disc}
+ {\txts\frac{1}{9}}\Gf{\rm b}{\rm b}_{\rm conn} + \dots
\phantom{\Bigg|}
\ee
As the notation indicates, for the heavy quarks we treat separately  quark-connected and disconnected contributions.
The dots stand for contributions that are too small to be of relevance in this work, namely disconnected diagrams
involving the bottom quark and the top contribution.
We use the corresponding notation for the different flavour contributions to $\ahvp\equiv\af{\gamma}{\gamma}$
as well as for  $\aSD\equiv \aSDf{\gamma}{\gamma}$, in particular 
\be \label{eq:defamuSD}
\aSDf{d}{e} = \Big(\frac{\alpha}{\pi m_\mu}\Big)^2 \int_0^\infty dx_0\; {\rm K}(m_\mu x_0)\; w_{\rm SD}(x_0) \,\Gf{d}{e}(x_0).
\ee

\subsection{General computational strategy \label{s:strat_sub}}

For a contribution to $\aSD$ of a given flavour,
we introduce the decomposition\footnote{With the standard choice of parameters of eq.\ (\ref{eq:stdwindparams}), $w_{\rm SD}(0)=0.995195$
and  $d w_{\rm SD}/dx_0(0)= -0.0125807$\,GeV.}
\ba\la{eq:amuSDfinalrep}
\aSDf{d}{e} &=& \aSDsubf{d}{e}(Q^2) + w_{\rm SD}(0)\, \bSDf{d}{e}(Q^2),
\ea
where
\ba
 \la{eq:defbSD}
\bSDf{d}{e}(Q^2) &=& \frac{16 \alpha^2 m_\mu^2}{9 Q^2} \Big(\Pif{d}{e}(Q^2) - \Pif{d}{e}(Q^2/4) \Big),
\\ \la{eq:defamuSDsub}
\aSDsubf{d}{e}(Q^2) &=& \Big(\frac{\alpha}{\pi m_\mu}\Big)^2 \int_0^\infty dx_0\; \Gf{d}{e}(x_0) \;{\rm K}_{\rm sub}^{\rm SD}(m_\mu,Q,x_0),
\\  \la{eq:defKsub}
{\rm K}_{\rm sub}^{\rm SD}(m_\mu,Q,x_0)&=&\Big(w_{\rm SD}(x_0) {\rm K}(m_\mu x_0)   - w_{\rm SD}(0)  \Big(\frac{16 \pi m_\mu^2}{3Q^2}\Big)^2 
\sin^4\Big(\frac{Q x_0}{4}\Big) \Big)   \,.
\ea
On the basis of eq.\ (\ref{eq:Kasympt}), the (leading)  $x_0^4$ part of the weight function defining $\aSD$
is cancelled in the integral of eq.\ (\ref{eq:defamuSDsub}). We expect this cancellation to reduce cutoff effects in the lattice calculation.
On the other hand, the subtracted quantity can be computed using the Adler function exclusively at virtualities greater or equal to $Q^2/4$.
The known good convergence properties of $D(Q^2)$ for $Q\gtrsim 2.5$\,GeV lead us to expect similarly good properties for
$\bSDf{d}{e}$.
In eq.\ (\ref{eq:defbSD}) applied to the isovector channel,  
we intend to compute $\bSDf{d}{e}$ by inserting the massless perturbative Adler function into eq.\ (\ref{eq:Del2}).
In other channels, the finite quark-mass effects in $\bSDf{d}{e}$ can be computed in lattice QCD up to the charm quark mass.

On the left hand side of Figure~\ref{fig:integrands} the kernel function 
${\rm K}_{\rm sub}^{\rm SD}(m_\mu,Q,x_0)$ of eq.~(\ref{eq:defKsub}) is shown
for several choices of the virtuality $Q$. The very-short distance region
of the integrand for $\aSD$ is excluded in all cases.
In the following, more specific aspects of our strategy are discussed for the
most important channels.

\subsubsection{The isovector contribution: test of massless perturbation theory}

We can write
\be\la{eq:aSDdiff}
\aSDsubf{d}{e}(\Qref^2)  - \aSDsubf{d}{e}(Q^2)
= w_{\rm SD}(0) \Big[ \bSDf{d}{e}(Q^2) - \bSDf{d}{e}(\Qref^2) \Big].
\ee
This equation can serve as a test to compare lattice results for the left-hand side to the perturbative Adler function
based calculation of the right-hand side. For orientation, at leading order in perturbation theory
\be
\bSDf{3}{3}(Q^2) = \frac{4\alpha^2 m_\mu^2}{9\pi^2 Q^2} \;  {\rm ln}(2)
\stackrel{Q=5\,{\rm GeV}}{=\!\!=\!\!=} 7.422\times 10^{-10} \qquad \textrm{(free massless quarks)}.
\ee

\subsubsection{Strategy for the $(u,d,s)$ isoscalar contribution}

Knowing $\aSDf{3}{3}$ from the lattice, one method of obtaining $\aSDf{8}{8}$ is via the decomposition
\be\la{eq:G88esti}
\aSDf{8}{8} =   \aSDf{3}{3} + \Delta_{\rm ls} \aSDO.
\ee
Only the term $\Delta_{\rm ls}\aSDO $ needs to be computed anew.
It is SU(3)$_{\rm f}$ breaking and therefore parametrically suppressed at short distances:
in the continuum, the linear combination of correlators $\Gf{8}{8} - \Gf{3}{3}$
has the leading parametric behaviour 
$\alpha_s(m_{\rm s}^2-m_{\rm l}^2)/|x_0|$ at $|x_0|\to0$,
thus making the corresponding integrand for $\aSD$ very suppressed at short distances.
Therefore no help from perturbation theory is needed to compute $\Delta_{\rm ls} \aSDO$.

Note that the estimator (\ref{eq:G88esti}) is entirely equivalent
to  proceeding via eq.\ (\ref{eq:amuSDfinalrep}), with  $\bSDf{8}{8}(Q^2)$ evaluated via
$ \bSDf{8}{8} = \bSDf{3}{3} + \Delta_{\rm ls}\bSD$ ,
where $\bSDf{3}{3}$ is taken from massless perturbation theory and $\Delta_{\rm ls}\bSD$ from the lattice.
However, it appears simpler to focus on  $\Delta_{\rm ls} \aSDO$,
since the latter quantity is expected to be much smaller than the already computed $\aSDf{3}{3}$.

\subsubsection{Strategy for the charm-connected contribution}

Similarly, the charm-connected contribution can be estimated according to eq.\ (\ref{eq:amuSDfinalrep}),
where the subtraction function $\bSDf{\rm c}{\rm c}(Q^2)$ is evaluated according to
\be
\bSDf{\rm c}{\rm c}_{\rm conn}(Q^2) = 2\bSDf{3}{3}(Q^2) + \Delta_{\rm lc}\bSD,
\ee
where the first term is taken from massless perturbation theory and $\Delta_{\rm lc}\bSD$
is evaluated on the lattice.

%% file: sec_lattice_setup.tex
\subsection{Lattice setup \label{s:lattice_setup}}
We refer to our previous publications~\cite{Gerardin:2019rua,Ce:2022kxy} 
for an in-depth overview of our
computational setup. Here, we only point out the refinements that we have
implemented in view of computing the short-distance contribution and $\amu$.

\input{./tables/tab_ens}

\subsubsection{Gauge ensembles}
We perform our computation on the $2+1$ flavour CLS ensembles with 
$\mathrm{O}(a)$ improved Wilson fermions and a tree-level $\mathrm{O}(a^2)$
improved Lüscher-Weisz gauge action \cite{Bruno:2014jqa, Bali:2016umi}.
Twisted-mass and RHMC determinant reweighting are used to stabilize the
simulations and to simulate the strange quark, respectively 
\cite{Clark:2006fx, Luscher:2012av, Mohler:2020txx, Kuberski:2023zky}.
Compared to our recent computation of the intermediate-distance contribution
in \cite{Ce:2022kxy}, we have extended the set of gauge ensembles that is
used in our study.
By including a second ensemble with physical quark masses we are
able to more tightly constrain mass-dependent cutoff effects which
otherwise could spoil our strategy to control the continuum extrapolation 
with ensembles at larger-than-physical light quark masses. 

As the sum of the bare sea quark masses is held constant for each value of
the bare coupling on the ensembles that have been used so far, 
the combination of meson masses
$m_K^2 + \textstyle{\frac{1}{2}}m_\pi^2$
is approximately constant along each chiral trajectory. To correct for a small
deviation from $m_K^{\rm phys}$ when approaching $m_\pi^{\rm phys}$, due to
cutoff effects and higher orders in chiral perturbation theory, we have
performed dedicated measurements to compute the derivatives of our observables
with respect to the quark masses in \cite{Ce:2022kxy}. This allowed us to 
constrain the dependence of $\aID$ on $m_K^2 + \textstyle{\frac{1}{2}}m_\pi^2$.
In view of the large statistical uncertainties of these derivatives when
considering the region of large Euclidean times, we have adapted our strategy 
by including four ensembles on a different chiral trajectory where the strange 
quark mass is approximately held at its physical value. 
Since the pion masses on these additional ensembles are in the range between
200\,MeV and 260\,MeV, the deviation from 
$(m_K^{\rm phys})^2 + \textstyle{\frac{1}{2}}(m^{\rm phys}_\pi)^2$
is at the level of a few percent on these ensembles.

An overview of the ensembles used in this work is given in table \ref{t:ensembles}.

\subsubsection{\texorpdfstring{$\mathrm{O}(a)$}{O(a)} improvement}

We perform full $\mathrm{O}(a)$ improvement of the action \cite{Bulava:2013cta}
and the observables used in this work. We use the local ($\loc$) and the
point-split ($\cons$) currents on the lattice, defined via
\begin{align}
J_{\mu}^{(\mloc),a}(x) &= \psib(x) \gamma_{\mu} T^a \psi(x) \,,\\
J_{\mu}^{(\mcons),a}(x) &= \frac{1}{2} \left(
\psib(x+a\hat{\mu})(1+\gamma_{\mu}) U^{\dag}_{\mu}(x)
T^a \psi(x) - \psib(x) (1-\gamma_{\mu} ) U_{\mu}(x)
T^a \psi(x+a\hat{\mu}) \right)  \,,
\label{eq:consvec1}
\end{align}
where $U_{\mu}(x)$ is the gauge link in the direction $\hat{\mu}$ associated
with site $x$. 
With the local tensor current defined as 
$\Sigma^{a}_{\mu\nu}(x) = -\frac{1}{2}\,
\overline{\psi}(x) [\gamma_{\mu}, \gamma_{\nu}] T^a \psi(x)$,
we perform chiral $\mathrm{O}(a)$ improvement of the currents via
\begin{equation}
\label{eq:impcv}
J^{(\alpha),a,I}_{\mu}(x) = J^{(\alpha),a}_{\mu}(x) +
a\cv^{(\alpha)}(g_0) \, \partial_{\nu} \Sigma^{a}_{\mu\nu}(x)
\,,\quad \alpha=\loc,\,\cons\,,
\end{equation}
with two independent sets of non-perturbatively determined coefficients
$\cv^{(\alpha)}$ from \cite{Gerardin:2018kpy} and \cite{Heitger:2020zaq}.
In past work, we have employed the symmetric discrete derivative
$\tilde{\partial}_{\nu} f(x) = (1/2a) \left( f(x+a) - f(x-a) \right)$ to 
compute the time derivative of the tensor current. At very short Euclidean
distances, where the vector-tensor correlator falls off steeply proportional
to $x_0^2$, the cutoff effects from the discrete derivative can be significant,
despite being $\mathrm{O}(a^2)$. By rewriting
\begin{align}
	\tilde{\partial} _{0} \Sigma^{a}_{\mu 0}(x) 
	\to
	\frac{1}{x_0^2} \left[
	\tilde{\partial} _{0} (x_0^2 \Sigma^{a}_{\mu 0}(x))
	- 2 x_0 \Sigma^{a}_{\mu 0}(x)
	\right]\,, \label{e:derivative}
\end{align}
the derivative acts on a function that varies more slowly and the size of 
the cutoff effects at short distances is reduced. Note that a similar effect
may be achieved by rewriting the TMR integral via integration by parts such
that the time derivative only acts on the functions ${\rm K}(m_\mu x_0)$ and $ w_{\rm SD}(x_0)$ which are defined in the continuum.
In this work, we use the derivative of eq.~(\ref{e:derivative}) to reduce the 
cutoff effects in the short-distance region.

\subsubsection{Tree-level improvement}
As pointed out in section \ref{s:strat_sub}, we reduce cutoff effects from 
very short Euclidean distances by evaluating a part of $\aSD$ 
perturbatively, thus cancelling potentially dangerous effects of
$\mathrm{O}(a^2\log(a))$. 
A further reduction of cutoff effects from short distances may be achieved
by computing these to tree-level in perturbation theory 
\cite{Ce:2021xgd, ExtendedTwistedMass:2022jpw} and subtracting them from the 
non-perturbatively calculated observable.
Denoting the tree-level evaluation of an observable $\mathcal{O}(a)$ by 
$\mathcal{O}^{\rm tl}(a)$, we can improve the approach to the continuum limit
by either of the replacements
\begin{align}
	\mathcal{O}(a) \rightarrow \mathcal{O}(a) \frac{\mathcal{O}^{\rm tl}(0)}{\mathcal{O}^{\rm tl}(a)}\,,\qquad \mathcal{O}(a) \rightarrow \mathcal{O}(a) - (\mathcal{O}^{\rm tl}(a) - \mathcal{O}^{\rm tl}(0))\,.
\end{align}
For $\aSDsub$, we find that
the multiplicative improvement seems to reduce the cutoff effects slightly 
better than the additive version and therefore use it in this work. 
Note that the additive improvement may be used to modify $\amu$ via
\begin{align}
	\amu(a) \rightarrow \amu(a) - [(\amu)^{\rm SD, tl}(a) - (\amu)^{\rm SD, tl}(0)]\,.
\end{align}

We use massless perturbation theory at leading order to compute 
$(\amu)^{\rm SD, tl}_{\rm sub}$.
To all orders in perturbation theory, the $\< \Sigma_{\mu\nu}(x) J_\lambda(0)\>$ correlator
is a pure O($a$) artefact in the massless theory.
Therefore, the currents are 
automatically $\mathrm{O}(a)$ improved, whereas the tree-level values 
$\cv^{\mloc, {\rm tl}}=0$ and $\cv^{\mcons, {\rm tl}}=0.5$ would have to be used for
massive correlators. We find that using the non-perturbatively determined values
for $\cv^{(\alpha)}$, the same ones as in the non-perturbative computation,
leads to a further reduction of cutoff effects.
We interpret this as a cancellation of effects of $\mathrm{O}(a^n), n\geq 2$
that are introduced by the derivative of the tensor current when 
performing $\mathrm{O}(a)$ improvement.

\subsubsection{Chiral-continuum fit}
Cutoff effects are clearly the biggest challenge for a precise computation
of $\aSD$ from lattice QCD. The reliable extrapolation of our results to the
continuum limit with a conservative estimate of the corresponding systematic
uncertainty is therefore the main challenge of this work. Even after applying
the techniques described above to tame discretization effects, higher order cutoff effects, compared to $a^2$ scaling,
 are visible in our data and thus have to be included in our fits. 
We note that it is not the relative size of the cutoff effects but this 
curvature that makes the extrapolation challenging.
Given the very precise data and the spread of quark masses on the ensembles
which are included in our computation, we can also resolve mass-dependent
cutoff effects. 

We therefore test a variety of ansätze to describe the dependence of 
our data on the lattice spacing and the quark masses. The fit to the
quark mass dependence is no particular challenge since we include two 
ensembles at physical quark masses and therefore do not need to extrapolate.
We find mild quark mass effects in the short-distance region.

We set the scale using the gradient flow scale $t_0/a^2$~\cite{Luscher:2010iy}
together with its physical value $\textstyle \sqrt{t_0^{\rm phys}}=0.1443(7)\,$fm which was determined 
in \cite{Strassberger:2021tsu} from a combination of $f_\pi$ and $f_K$.
As an alternative, we have in the past directly set the scale with the pion decay
constant by the ratio $f_\pi^{\rm phys} / (a f_\pi)$. 
This was found to reduce the slope of both chiral and continuum extrapolations 
for $\ahvp$ in \cite{Gerardin:2019rua}. Since the short-distance contribution
has only a mild dependence on $m_\pi$ that is significantly different from 
that of $f_\pi$, we do not use $f_\pi$-rescaling in this work. 
However, we have verified that we would obtain very similar results to 
the ones quoted in this work, albeit with larger uncertainties due to the
enhanced difficulty in the extrapolation of the data to the physical point.

We define our scheme for isospin-symmetric QCD via the conditions
\begin{eqnarray}
m_\pi = (m_{\pi^0})_{\rm phys}, \qquad 
2 m^2_K - m^2_{\pi} = (m^2_{K^+} + m^2_{K^0}  - m^2_{\pi^+})_{\rm phys},
\end{eqnarray}
corresponding to
\begin{eqnarray}
m_\pi = 134.9768(5)~\MeV \,, \quad m_K = 495.011(10)~\MeV \,.
\end{eqnarray}
We also introduce the dimensionless combinations
\begin{equation}
\Phi_2 = 8 t_0 m_{\pi}^2,\qquad
\Phi_4 = 8 t_0 ( m_K^2 + {\textstyle\frac{1}{2}} m_{\pi}^2  )\,.
\end{equation}

To describe the chiral dependence of the observables $\mathcal{O}$ computed in 
this work, we always include a term that is linear in $m_\pi^2$ and allow for
another term that includes a higher order correction, 
\begin{align} \label{e:cc_pi}
	\mathcal{O}(\Xpi) =&{}  \mathcal{O}(\Xpi^{\rm phys}) + \gamma_1 \, \left( \Xpi - \Xpi^{\rm phys} \right)  +
	\gamma_2 \left( f_{\rm ch}(\Xpi) -  f_{\rm ch}(\Xpi^{\rm phys})\right)\\
	&\text{where } f_{\rm ch} \in  \{ 
	\Xpi\log (\Xpi)\,;\enspace
	\Xpi^2\,
	\} \nonumber\,.
\end{align}
The strange quark mass dependence is parametrized via the parameter
$\XK \propto m_K^2 + \frac{1}{2} m_\pi^2$ which is 
close to its physical value on all of the ensembles considered in this work. 
We fit to
\begin{align}\label{e:cc_K}
	\mathcal{O}(\XK) =&{}  \mathcal{O}(\XK^{\rm phys}) + \gamma_0 \, \left( \XK - \XK^{\rm phys} \right)\,.
\end{align}

To describe the cutoff effects with sufficient quality, we need to allow for
a number of terms, guided by Symanzik effective theory. Our most general
ansatz is
\begin{align}\label{e:cc_a}
	\mathcal{O}(\Xa) =&{} \beta_2 \, \Xa^2 + \beta_3 \, \Xa^3  + \beta_4 \, \Xa^4 + \delta_2 \, \Xa^2 \left( \Xpi - \Xpi^{\rm phys} \right)
	\\\nonumber
	&{}+ \delta_3 \, \Xa^3 \left( \Xpi - \Xpi^{\rm phys} \right)
	+ \epsilon_2 \, \Xa^2 \left( \XK - \XK^{\rm phys} \right)\,,
\end{align}
where $\Xa = \sqrt{a^2 / (8t_0)}$. It is not possible to constrain all fit
parameters at once.
We build a variety of different descriptions of the chiral and cutoff effects 
by setting one or several of the parameters $\gamma_i$, $\beta_i$, $\delta_i$,
$\epsilon_2$ to zero.

We note that, in an $\mathrm{O}(a)$ improved theory, the prediction from 
Symanzik effective theory for the scaling is 
$a^2 \left[\alpha_{\rm s} (1/a)\right]^{\hat{\Gamma}}$ 
with $\hat{\Gamma}$ the leading anomalous dimension 
\cite{Husung:2019ytz,Husung:2021mfl}. 
Any curvature in $a^2$ may therefore be due to the modified scaling 
behaviour and not due to higher order cutoff effects.%
\footnote{Note that this is not related to the log-enhanced cutoff effect 
that we have eliminated from our calculation by the
changes to the TMR kernel function.}
The leading anomalous dimension for local quark bilinears with vector quantum
numbers has been computed to be $\hat{\Gamma} = 0$ in the $\mathrm{O}(a)$
improved action used in this work \cite{Husung:Lattice2023}. 
The lowest anomalous dimension
from the gauge action is $0.76$ \cite{Husung:2021mfl} and larger than the
smallest non-zero contribution from the current, which is $0.395$.
We allow for effects from beyond leading order terms by including the
modification
\begin{align}
	\Xa^2 \rightarrow [\alpha_{\rm s}(1/a)]^{0.395} \Xa^2
\end{align} 
in our continuum extrapolations, thereby doubling the number of fit ansätze.
We use five-loop running \cite{Baikov:2016tgj}, starting at  
$\Lambda^{(3)}_{\overline{\rm MS}}$ from \cite{Bruno:2017gxd}, to determine
the running-coupling constant.

Generally, we find that fits with $\hat{\Gamma}=0$ are favoured over fits with
$\hat{\Gamma}=0.395$. However, the latter have non-vanishing weights in our
averages and prefer continuum extrapolated results that are shifted to slightly
smaller values. The effect of including a non-vanishing anomalous dimension 
is found to be significantly smaller than the overall uncertainty in all cases
considered in this work.

We extend our explorations of the parameter space for the chiral and continuum
extrapolations by performing cuts in the data and repeating the analysis with 
reduced data sets. 
We perform cuts in the lattice spacing by removing the coarsest or
the two coarsest lattice spacings from our analysis. Furthermore, we perform
cuts by excluding all ensembles with $m_\pi < 400\,$MeV, or with 
$m_\pi < 300\,$MeV respectively in the case of contributions that vanish at the 
SU(3) symmetric point, from our analysis.

To assign and compare fit qualities in view of the change of the number of fit parameters
and data points, we apply the Akaike information criterion \cite{Akaike1998}
and the model averaging method from \cite{Jay:2020jkz}.
We assign a weight 
\begin{equation}
\label{eq:Akweight}
w_i = N \exp\left[ - \frac{1}{2} \left( \chi_i^2 + 2 k_i - 2 n_i \right) \right]
\end{equation}
to each fit, where $k_i$ is the number of fit parameters and $n_i$ is the number of data points in the fit with minimized $\chi_i^2$.
The normalization $N$ is such that $\sum_i w_i = 1$.%
\footnote{See the discussion in \cite{Borsanyi:2020mff,Neil:2023pgt} concerning slightly modified weights in 
the presence of cuts in the data and the differences between them. We find no
significant difference in our results when using one of these modified weights.}

We obtain the central value and statistical uncertainty of an observable
$\mathcal{O}$ by a weighted average over all analyses
\begin{equation}
\bar{\mathcal{O}} = \sum_i w_i \mathcal{O}_i \,.
\label{eq:akaike_mean}
\end{equation}
Our estimate of the systematic error associated with the extrapolation to the physical point is given by
\begin{equation}
(\delta \mathcal{O})_{\rm syst}^2 = \sum_i w_i (\mathcal{O}_i - \bar{\mathcal{O}})^2 \,.
\label{eq:akaike_syst}
\end{equation}
Statistical uncertainties are determined and propagated using the 
$\Gamma$-method in the implementation of the \texttt{pyerrors} 
package \cite{Wolff:2003sm,Ramos:2018vgu,Joswig:2022qfe}.

%% file: tables/tab_ens.tex
\begin{table}[!htbp]
\caption{Parameters of the simulations: the bare coupling $\beta = 6/g_0^2$, the temporal boundary conditions, open (o) or anti-periodic (p), the lattice dimensions, the lattice spacing $a$ in physical units based on \cite{Strassberger:2021tsu, RQCD:2022xux}, the pion and kaon masses, the physical size of the lattice and the length of the Monte Carlo chain in Molecular Dynamics Units (MDU). Ensembles with an asterisk are not included in the final analysis but used to control finite-size effects. Ensembles with a dagger are only on the chiral trajectory where $m_{\rm s} \approx m_{\rm s}^{\rm phys}$. Ensemble N452, marked with a plus, is only used in the computation of isospin breaking effects.
}
\vskip 0.1in
\begin{tabular}{lcl@{\hskip 01em}c@{\hskip 02em}c@{\hskip 02em}c@{\hskip 01em}c@{\hskip 01em}c@{\hskip 01em}c@{\hskip 01em}c@{\hskip 01em}
	}
\hline
Id   & $\quad\beta\phantom{\Big|}\quad$   & bc & $\textstyle\big(\frac{L}{a}\big)^3\times\frac{T}{a}$   & $a\,[{\rm fm}]$   & $m_\pi\,[{\rm MeV}]$   & $m_K\,[{\rm MeV}]$   &   $m_\pi L$ &   $L\,[{\rm fm}]$ 
& MDU
\\
\hline
A653 & 3.34 & p  & $24^3 \times 48$  & 0.0972(10) & 430(5) & 430(5) & 5.1 & 2.3 & 20200 \\
A654 &      & p  & $24^3 \times 48$  &            & 338(5) & 462(6) & 4.0 & 2.3 & 16000 \\
\hline
H101 & 3.4  & o  & $32^3 \times 96$  & 0.0849(9)  & 424(5) & 424(5) & 5.8 & 2.7 &  8064 \\
H102 &      & o  & $32^3 \times 96$  &            & 358(5) & 445(5) & 4.9 & 2.7 &  7832 \\
H105$^{*}$ &      & o  & $32^3 \times 96$  &            & 283(4) & 470(6) & 3.9 & 2.7 &  8260 \\
N101 &      & o  & $48^3 \times 128$ &            & 282(4) & 468(5) & 5.8 & 4.1 &  6376 \\
C101 &      & o  & $48^3 \times 96$  &            & 222(3) & 478(5) & 4.6 & 4.1 &  8000 \\
C102$^{\dag}$ &      & o  & $48^3 \times 96$  &            & 225(3) & 506(6) & 4.6 & 4.1 &  6000 \\
D150 &      & p  & $64^3 \times 128$ &            & 131(3) & 484(6) & 3.6 & 5.4 &  1616 \\
\hline
B450 & 3.46 & p  & $32^3 \times 64$  & 0.0751(8)  & 422(5) & 422(5) & 5.1 & 2.4 &  6448 \\
S400 &      & o  & $32^3 \times 128$ &            & 355(4) & 447(5) & 4.3 & 2.4 & 11492 \\
N452$^{+}$ &      & p  & $48^3 \times 128$ &            & 356(4) & 447(5) & 6.5 & 3.6 &  4000 \\
N451 &      & p  & $48^3 \times 128$ &            & 291(4) & 468(5) & 5.3 & 3.6 &  4044 \\
D450 &      & p  & $64^3 \times 128$ &            & 219(3) & 483(5) & 5.3 & 4.8 &  2000 \\
D451$^{\dag}$ &      & p  & $64^3 \times 128$ &            & 220(3) & 510(6) & 5.3 & 4.8 &  3700 \\
D452 &      & p  & $64^3 \times 128$ &            & 156(3) & 490(6) & 3.8 & 4.8 &  4000 \\
\hline
H200$^{*}$ & 3.55 & o  & $32^3 \times 96$  & 0.0635(6)  & 423(5) & 423(5) & 4.4 & 2.0 &  8000 \\
N202 &      & o  & $48^3 \times 128$ &            & 418(5) & 418(5) & 6.5 & 3.0 &  3200 \\
N203 &      & o  & $48^3 \times 128$ &            & 349(4) & 447(5) & 5.4 & 3.0 &  6172 \\
N200 &      & o  & $48^3 \times 128$ &            & 286(4) & 468(5) & 4.4 & 3.0 &  6848 \\
D251 &      & p  & $64^3 \times 128$ &            & 286(3) & 467(5) & 5.9 & 4.1 &  5968 \\
D200 &      & o  & $64^3 \times 128$ &            & 202(3) & 486(5) & 4.2 & 4.1 &  8004 \\
D201$^{\dag}$ &      & o  & $64^3 \times 128$ &            & 202(3) & 507(6) & 4.2 & 4.1 &  4312 \\
E250 &      & p  & $96^3 \times 192$ &            & 132(2) & 495(6) & 4.1 & 6.1 &  4640 \\
\hline
N300 & 3.7  & o  & $48^3 \times 128$ & 0.0491(5)  & 425(5) & 425(5) & 5.1 & 2.4 &  8188 \\
N302 &      & o  & $48^3 \times 128$ &            & 350(5) & 456(6) & 4.2 & 2.4 &  8804 \\
J303 &      & o  & $64^3 \times 192$ &            & 260(3) & 480(5) & 4.1 & 3.1 &  8584 \\
J304$^{\dag}$ &      & o  & $64^3 \times 192$ &            & 263(3) & 530(6) & 4.2 & 3.1 &  6508 \\
E300 &      & o  & $96^3 \times 192$ &            & 177(2) & 497(6) & 4.2 & 4.7 &  4548 \\
\hline
J500 & 3.85 & o  & $64^3 \times 192$ & 0.0386(4)  & 417(5) & 417(5) & 5.2 & 2.5 & 15000 \\
J501 &      & o  & $64^3 \times 192$ &            & 337(4) & 450(5) & 4.2 & 2.5 & 15680 \\
\hline
 \end{tabular} 
\label{t:ensembles}
\end{table}

%% file: sec_results.tex
\section{Results \label{s:results}}

\begin{figure}[t]
	\includegraphics*[width=0.48\linewidth]{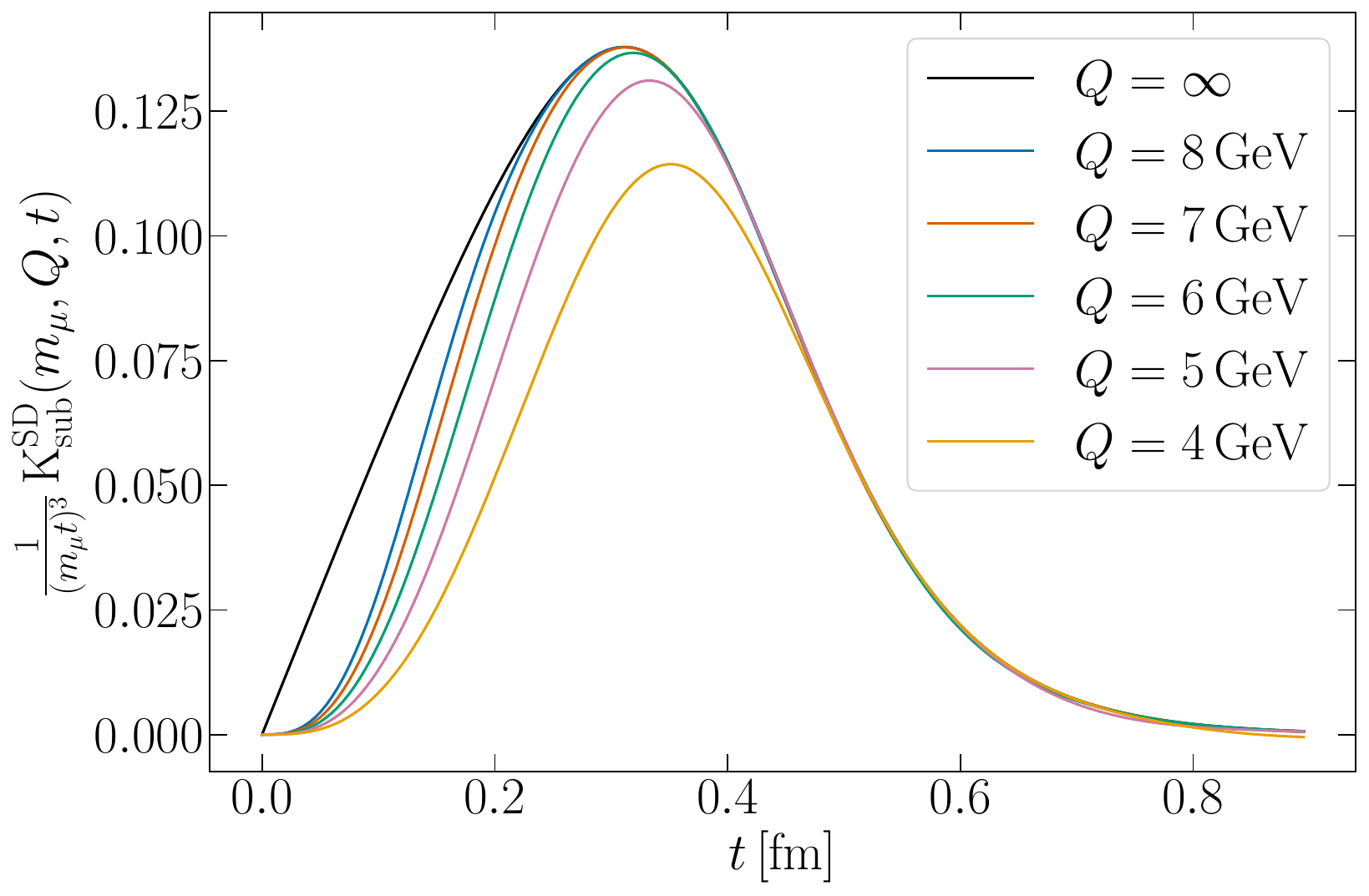}%
	\hspace{.01\linewidth}%
	\includegraphics*[width=0.50\linewidth]{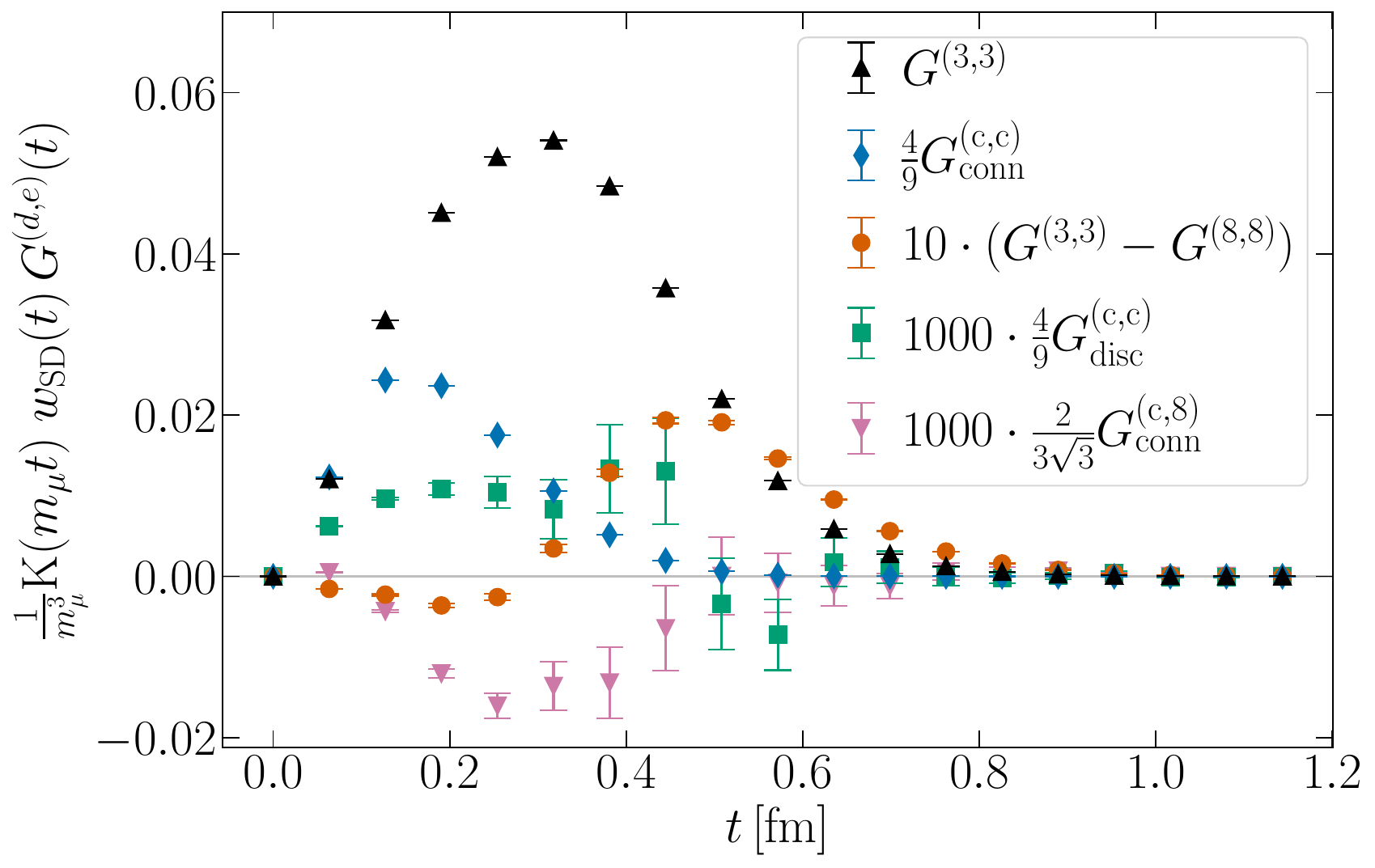}%
	\caption{Left: Illustration of the modified short-distance kernels according
		to eq.~(\ref{eq:defamuSDsub}) with varying virtuality.
	        Right: Integrands for the computation of $\aSD$ on ensemble E250 with near physical quark masses at $a\approx 0.064\,$fm. Smaller contributions have been scaled as indicated in the plot.
	}
	\label{fig:integrands}
\end{figure}

To illustrate the behaviour and the relative size of the contributions to 
$\aSD$, we show in the right panel of Figure~\ref{fig:integrands} integrands 
of the various contributions according to eq.~(\ref{eq:defamuSD}), i.e., without
subtraction, where the smaller contributions are scaled as indicated 
in the figure to show them on the same scale. The $\loc\cons$ discretization
is displayed for the quark-connected contributions and the $\cons\cons$ 
discretization for the two quark-disconnected contributions.
The charm-connected correlation function contributes significantly to the 
final result for $\aSD$. Except for the two charm-disconnected contributions,
the relative statistical uncertainties are very small. 

As outlined above, we non-perturbatively compute the subtracted observables
$\aSDsubf{d}{e}(Q^2)$ according to eq.~(\ref{eq:defamuSDsub}). 
After the combination with $\bSDf{d}{e}(Q^2)$, eq.~(\ref{eq:defbSD}), 
the dependence on the virtuality $Q$ has to cancel. 
We perform our calculation for several values of $Q$ to explicitly test this.
For our final result, we choose $Q=5\,$GeV as a good compromise between 
a high enough scale for the perturbative evaluation of the Adler function
and a smoothly varying kernel function 
${\rm K}_{\rm sub}^{\rm SD}(m_\mu,Q,x_0)$ on the lattice.

If not stated explicitly, all results for the quantities $\bSDf{d}{e}$, $\aSDsubf{d}{e}$ and $\aSD$ are from here on
expressed in units of $10^{-10}$.

\subsection{Perturbative evaluation of $\bSD(Q^2)$ in the isovector channel}

We have used the perturbative coefficients of the Adler function as
given in~\cite{Baikov:2010je} up to $\alpha_s^3$ and taking the
O($\alpha_s^4$) term from~\cite{Baikov:2008jh}. The effect is to multiply the parton-level
prediction by the factor $(1+ (\alpha_s/\pi) + \sum_{n=2}^4 (\alpha_s/\pi)^n \,d_n )$.

We use the perturbative Adler function in the $N_f=3$ massless theory 
using $\Lambda_{\overline{\rm MS}}=0.338(12)\,$GeV~\cite{FlavourLatticeAveragingGroupFLAG:2021npn}.
The QCD beta function was taken into account up to three-loop order included,
and we used the corresponding large-$Q^2$ asymptotic solution for $\alpha_s$ throughout,
after having checked that using instead the numerical solution for $\alpha_s$ makes an entirely negligible difference.
As expected, the convergence of perturbation theory is excellent. For instance, at $Q=5\,$GeV,
the set of values of the subtraction term, from the parton-level prediction to the highest-order prediction read 
\be\la{eq:b33pertNf3}
\bSDf{3}{3}(25\,{\rm GeV}^2)
= \{7.422,~7.971,~8.037,~8.057,~8.068 \}.
\ee
Table \ref{t:b_pert} lists the perturbative results for several values of $Q^2$.
Secondly, the uncertainty of $\Lambda_{\overline{\rm MS}}$ induces an absolute uncertainty of $0.011$ on $\bSDf{3}{3}(25\,{\rm GeV}^2)$.
Thirdly, the sea-quark effect due to the strange quark is small. In the estimates above, the strange-quark mass is treated as zero.
Evaluating the expression in the $N_f=2$ massless theory (corresponding to $m_s=\infty$)
with $\Lambda_{\overline{\rm MS}}=0.330\,$GeV~\cite{FlavourLatticeAveragingGroupFLAG:2021npn}, we obtain
\be\la{eq:b33pertNf2}
\bSDf{3}{3}(25\,{\rm GeV}^2) = 8.020 \quad\qquad (N_f=2).
\ee
The comparison of Eqs.\ (\ref{eq:b33pertNf3}) and (\ref{eq:b33pertNf2})
shows that the effect of changing $m_s=0$ to $m_s\simeq 100$\,MeV must be negligibly small,
since it is parametrically suppressed by $m_s^2/Q^2$.
Finally, we have checked that the phenomenologically estimated gluon and quark `condensates'~\cite{Eidelman:1998vc}
make a negligible contribution.

Anticipating that the quantity $\bSDf{3}{3}(25\,{\rm GeV}^2)$ is the only perturbative prediction that
enters our final result for $\aSD$, we take as its full uncertainty the linear sum of the error from $\Lambda_{\overline{\rm MS}}$
and the size of the last (O($\alpha_s^4$)) perturbative term,
\be
\bSDf{3}{3}(25\,{\rm GeV}^2) = 8.068 \pm 0.022.
\ee

\begin{table}[!t]
	\centerline{\begin{tabular}{|r|@{$\qquad$}r|@{$\qquad$}r|}
			\hline
			$Q$[GeV]    & $ \bSDf{3}{3}(Q^2) $ & $ \bSDf{\rm c}{\rm c}(Q^2) $ \\
			\hline
			3.5 &   16.740(43)(34) & 20(12) \\
			4.0 &   12.727(26)(22) & 16.3(5.1) \\ 
			5.0 &  8.068(11)(11)  &  11.4(1.3) \\
			6.0 &   5.567(6)(6) & 8.57(45) \\ 
			7.0 &   4.072(3)(4) & 6.66(17) \\ 
			8.0 &  3.106(2)(3)  & 5.31(8) \\ 
			\hline
	\end{tabular}}
	\caption{The perturbatively computed subtraction terms $ \bSDf{3}{3}(Q^2) $  and $ \bSDf{\rm c}{\rm c}(Q^2) $ (in units of $10^{-10}$)
		for the isovector and the charm correlator respectively.
		The former is computed using the $N_f=3$ massless perturbative Adler function to O$(\alpha_s^4)$ included, the latter
		with the massive perturbation theory to O$(\alpha_s^2)$ included, in the expansion with respect to the ratio
		$(m_c^2/Q^2)$ to third order included. The first number in brackets indicates the absolute size of the last O($\alpha_s^n$) term
		included in the estimate and gives an indication of the  uncertainty related to the convergence of the perturbative series.
                For $ \bSDf{3}{3}$, the uncertainty induced by that on $\Lambda_{\overline{\rm MS}}$ is indicated as well.\label{t:b_pert}}
\end{table}

\subsection{Non-perturbative evaluation of $\aSDsub(Q^2)$ in the isovector channel}

\begin{figure}[t]
	\includegraphics*[width=0.49\linewidth]{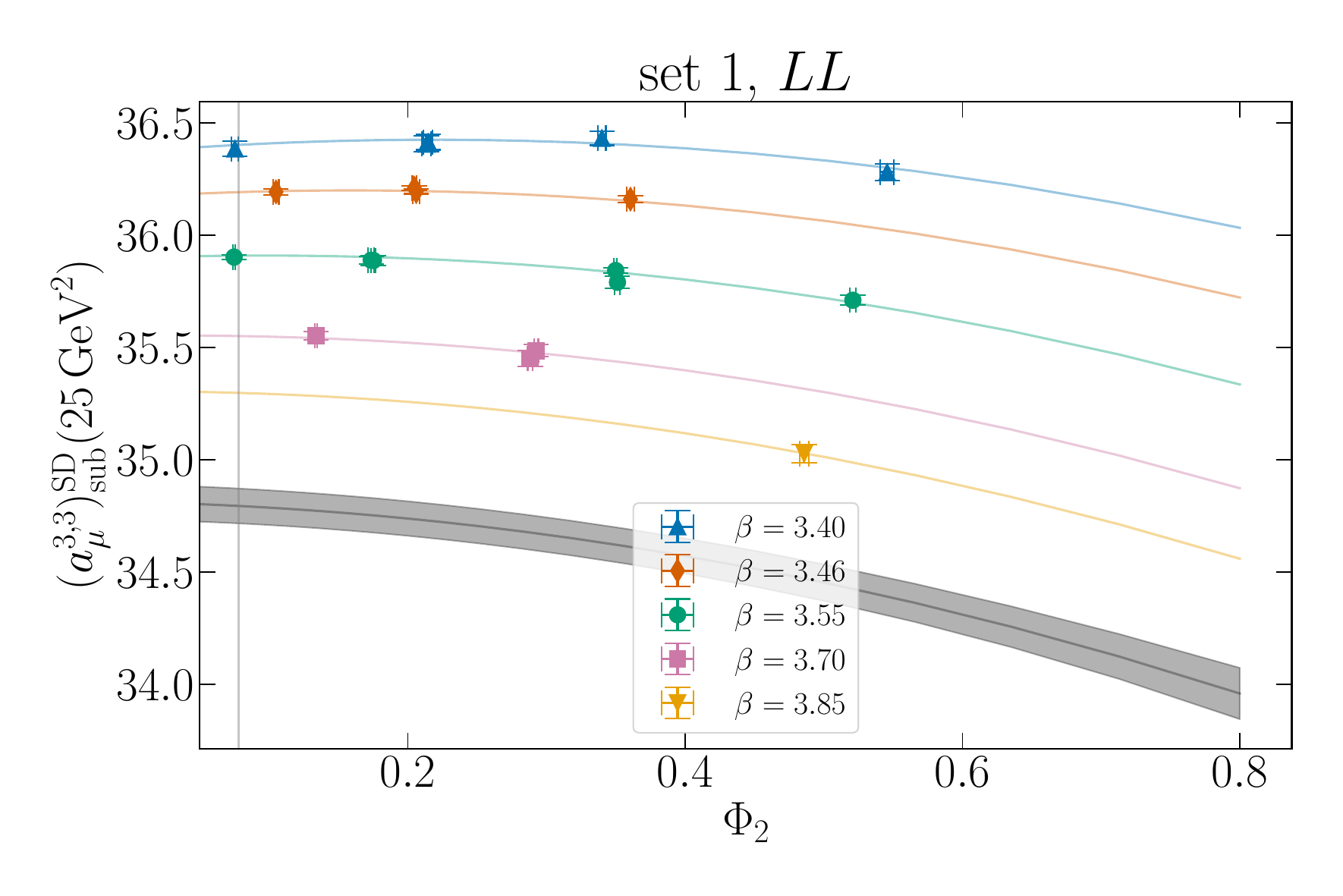}%
	\includegraphics*[width=0.49\linewidth]{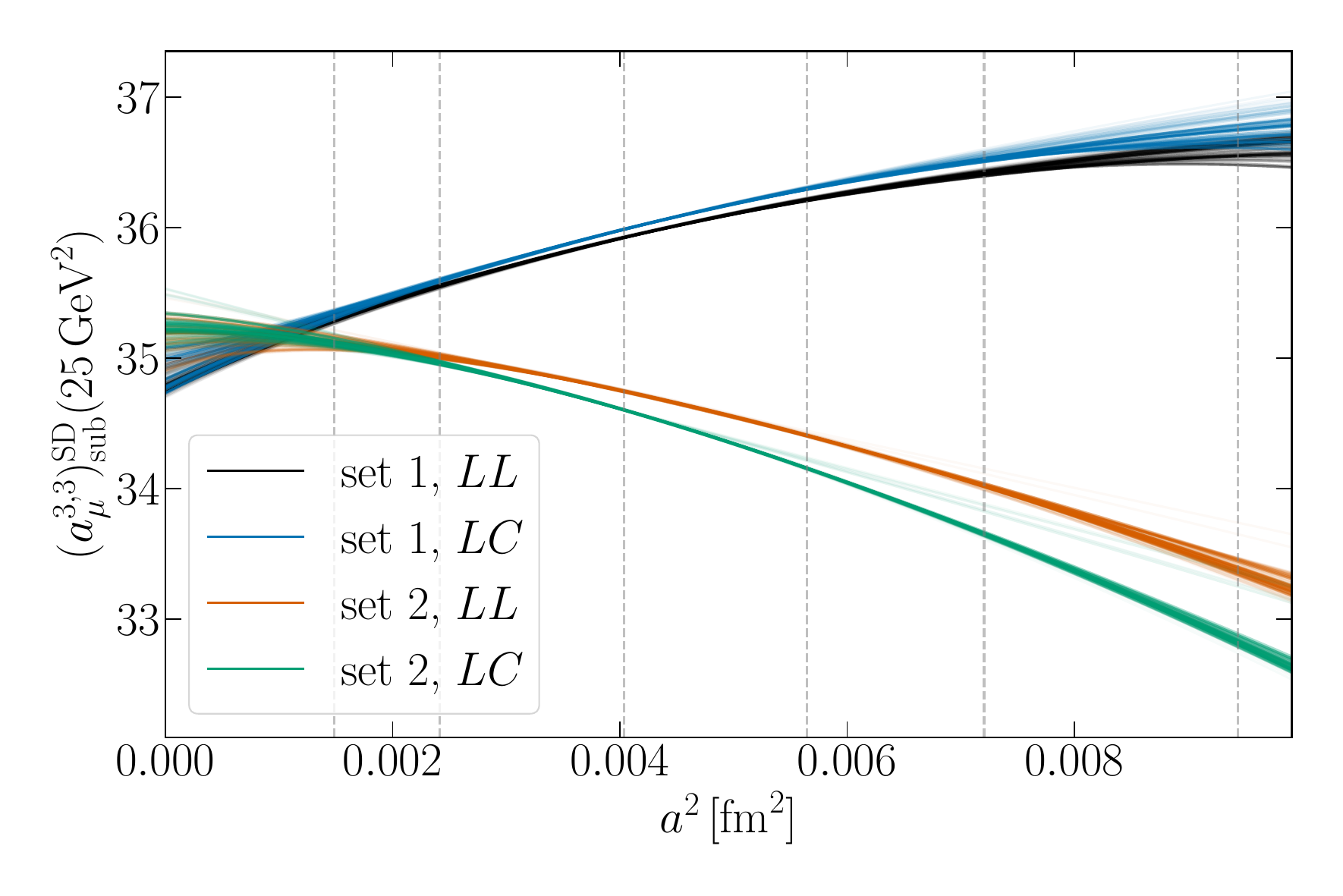}%
	\caption{
	Illustration of fits to $\aSDsubf{3}{3}(25\,{\rm GeV}^2)$.
	Left: Best fit, according to its model weight, to the data for set 1, 
	$\loc\loc$. The data is corrected for deviation from physical $\Phi_4$.
	The coloured lines show the evaluation of the fit function at finite 
	lattice spacing. The black line, together with the grey band, show the
	dependence on $\Phi_2$ in the continuum limit.
	Right:  
	Approaches to the continuum limit for four sets of data based on
	the improvement schemes of set 1 and 2 and the $\loc\loc$ and $\loc\cons$
	discretizations of the current based on a scan over fit models for
	$\aSDsubf{3}{3}(25\,{\rm GeV}^2)$. 
	Each line shows the result from one single fit and the opacity of 
	the lines corresponds to the weight of the fit in the model average. 
	Dashed vertical lines indicate the lattice spacings used in this work.
	}
	\label{f:aSDI1_fits}
\end{figure}

In the short-distance regime, no signal-to noise problem hinders the computation
in the isovector channel and the precise lattice data can be integrated 
by summation.
We evaluate $\aSDsub$ at several values of $Q^2$ to be able to monitor the
convergence of the sum in eq.~(\ref{eq:amuSDfinalrep}) and to investigate 
whether the choice of $Q^2$ has an influence on the systematic uncertainties
of the lattice evaluation.

The small finite-volume effects are corrected using the method by Hansen 
and Patella \cite{Hansen:2019rbh,Hansen:2020whp}, see \cite{Ce:2022kxy} for 
the details of our implementation. We find the size of the correction
to be at most two permille on the ensembles included in our computation.
Finite-volume corrections due to kaon loops are non-negligible close to the
SU(3) symmetric point, where pion and kaon masses are of similar order. 
We use the same procedure as for the effects from pions to estimate the 
corresponding correction. Both contributions and their sum are given 
in table~\ref{tab:fvc}.
A flat $25\%$ systematic uncertainty is assigned to the corrections.

As explained above, we scan a variety of ansätze for the chiral-continuum
extrapolation to determine $\aSDsubf{3}{3}$ at the physical point.
The best fit according to the AIC for the data of set 1 with the $\loc\loc$ discretization is shown on the left-hand side 
of Figure~\ref{f:aSDI1_fits}. 
The minimized $\chi^2$ is $7.64$ for 11 degrees of freedom (dof).
We show the dependence of
$\aSDsubf{3}{3}(25\,{\rm GeV}^2)$ on the light quark
mass proxy $\Phi_2$, where the data has been corrected for deviations
from $\Phi_4^{\rm phys}$ and cuts to $m_\pi < 400\,{\rm MeV}$ and $a < 0.09\,$fm have been applied. The data and the fit function are shown for the five values of the inverse gauge coupling that are included in this fit. 
As visible from the figure, the quark mass effects are mild but depend on
the lattice spacing. The fit also includes cutoff effects proportional
to $a^2\Phi_4$ which are not visible in the plot.

On the right-hand side of Figure~\ref{f:aSDI1_fits}, we show the approach to
the continuum limit at physical quark masses, as determined from the scan over 
all fit models for both discretizations of the current and both 
improvement schemes. Each line corresponds to one fit and the opacity 
is related to the model weight of the fit according to the AIC.
The relative size of the cutoff effects, about $10\%$ in the most extreme
case, is not larger than in our computation of the intermediate-distance
window observable \cite{Ce:2022kxy}, thanks to some of the improvements 
described in section \ref{s:lattice_setup}. For the same reason, the 
difference between the two discretizations of the current is small.

The two sets of improvement schemes lead to cutoff effects that have opposite signs. A similar behaviour has been
observed for the case of the intermediate-distance window, where the 
extrapolations of both data sets agreed in the continuum limit, 
taking into account the curvature in the extrapolation of the set~2 data. 
In the case of $\aSDsubf{3}{3}$ we observe that most of the fits from set~2
with a large model weight favour slightly larger values than the fits to 
the data of set~1. However, for some fits we observe a strong curvature 
at very small values of the lattice spacing such that they are compatible 
with the more benign extrapolations for set 1.
In the current situation, where we do not have more information on the 
behaviour towards the continuum limit, e.g., by adding data at even 
finer lattice spacings, we decide to take the discrepancy between the two
data sets into account as systematic uncertainty of our continuum extrapolation.

We perform the model average over the four data sets, using a flat relative 
weight between the sets and arrive at 
\begin{align}\label{e:result_aSDsub33}
	\aSDsubf{3}{3}(25\,{\rm GeV}^2) = 
	35.03(4)_{\rm stat}(21)_{\rm syst}(3)_{\rm scale}[22] 
	\,.
\end{align}
Whereas the systematic uncertainty of $0.21$ amounts to
about $0.6\%$ of the value of $\aSDsubf{3}{3}$, we note that it is
sub-permille with respect to $\af{3}{3}$. 
The description of the short-distance cutoff effects therefore does not seem to be a roadblock
towards the target precision for the full HVP.

The qualitative behaviour of the continuum extrapolation does not vary with 
the choice of $Q^2$. A smaller value of $Q^2$, which removes more of the 
potentially difficult ultra short-distance region, does not lead to a better
agreement of the results based on the two improvement schemes. 

\subsection{Combination of perturbative and non-perturbative evaluations in the isovector channel}

\begin{figure}[t]
	\includegraphics*[width=0.48\linewidth]{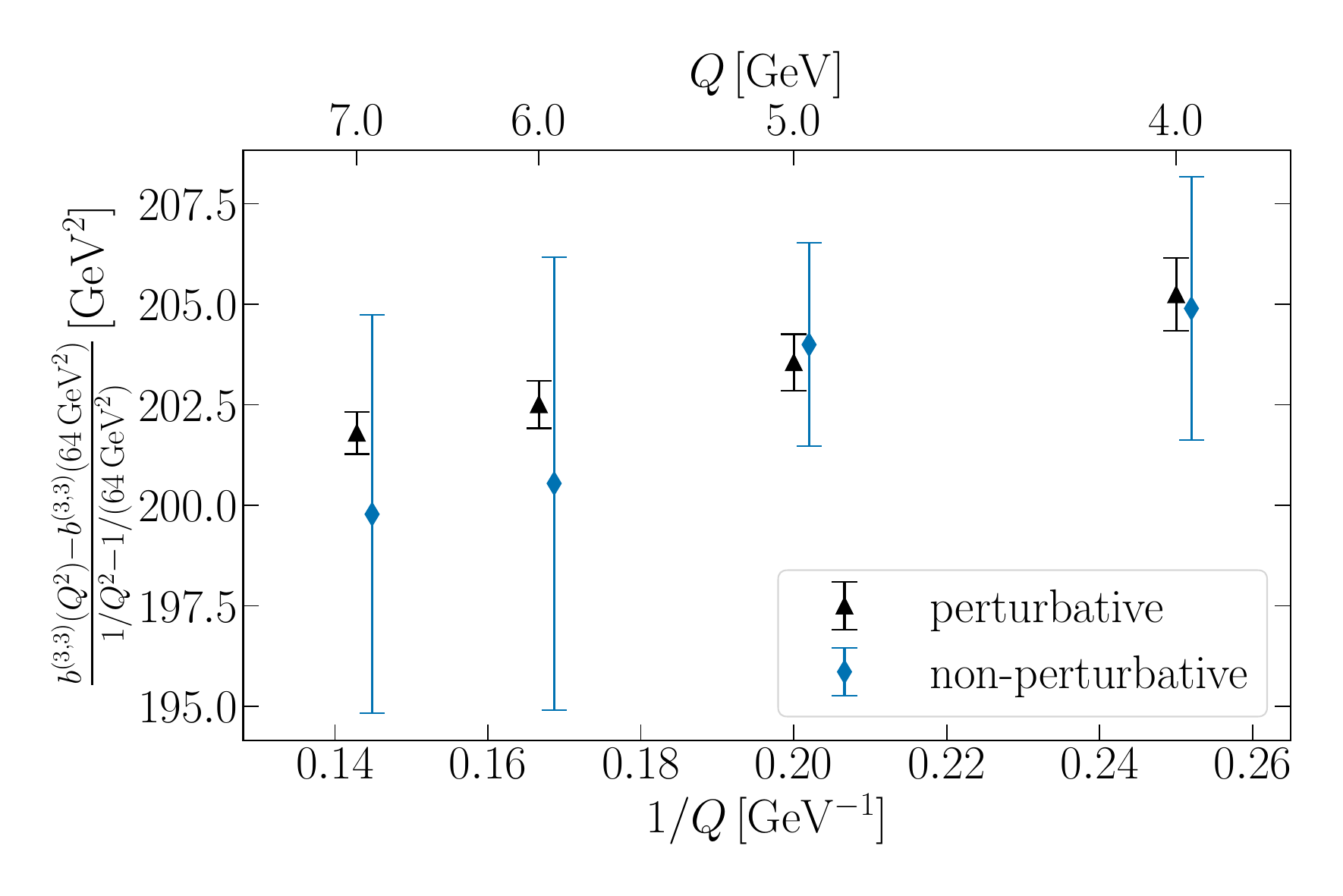} \
	\includegraphics*[width=0.48\linewidth]{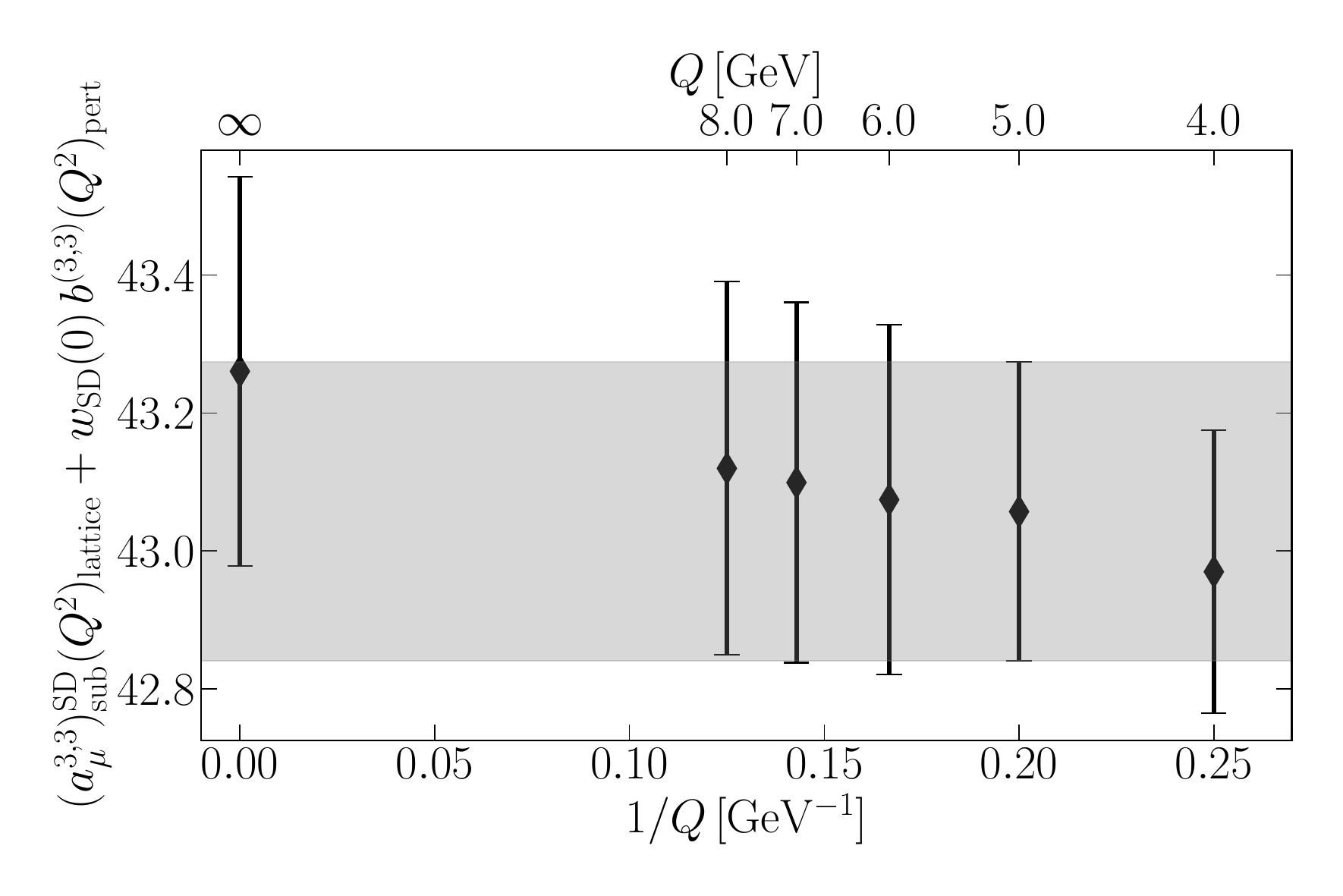} 
	
	\caption{Left: 
		The ratio $(\bSDf{3}{3}(Q^2) - \bSDf{3}{3}(\Qref^2)) / (1/Q^2 - 1/\Qref^2)$ with $\Qref^2=(8\,{\rm GeV})^2$ 
		for the isovector channel, computed non-perturbatively in lattice QCD
		vs.\ using the perturbative Adler function (see eqs.\ (\ref{eq:Del2}), (\ref{eq:defbSD}) and (\ref{eq:aSDdiff})). For visibility, the data is slightly displaced horizontally.
		Right: Stability of the result for $\aSDf{3}{3}$ as a function of the  virtuality $Q^2$ chosen in the 
		perturbatively treated subtraction term. Choosing $Q=\infty$ corresponds to the entire $\aSDf{3}{3}$ being computed in lattice QCD
		without any use of perturbation theory. }
	\label{fig:I1_combination}
\end{figure}

The non-perturbative evaluation of $\aSDsub$ at several values of $Q^2$ gives
us the possibility to monitor the convergence of perturbation theory and
to directly compare the perturbative and non-perturbative evaluations of
the difference in eq.~(\ref{eq:aSDdiff}). We choose the reference energy 
$\Qref = 8\,$GeV where the convergence of perturbation theory is expected
to be excellent.

In the non-perturbative evaluation, we compute the left hand side of
eq.~(\ref{eq:aSDdiff}).
We get compatible results when performing the difference either in the 
continuum limit or when subtracting ensemble by ensemble and extrapolating
the difference.
We choose the second version for the comparison with perturbation theory
in Figure ~\ref{fig:I1_combination}. It can be seen that both approaches give
essentially the same results down to $Q=4\,$GeV, confirming the good
convergence of the perturbative series.

On the right-hand side of Figure~\ref{fig:I1_combination} we depict the final 
quantity $\aSDf{3}{3}$ depending on the choice of $Q$. Any dependence on $Q$ should be removed after combining lattice and perturbation theory and indeed all results are entirely compatible with each other. For our final estimate, we use $Q=5\,$GeV and obtain
\begin{align}
\label{e:result_aSD33}
	\aSDf{3}{3} 
	= 
	43.06(4)_{\rm stat}(21)_{\rm syst}(3)_{\rm scale}[22]
	\,.
\end{align} 

We also show a result for the direct evaluation of $\aSDf{3}{3}$, without 
any subtraction in the kernel function, in Figure~\ref{fig:I1_combination},
where we have ignored the possibility of cutoff effects of
$\mathrm{O}(a^2\log(a))$~\cite{Ce:2021xgd} being present in the lattice data
and employed the same set of continuum extrapolations
as for the subtracted quantity. The agreement with our final result is an 
indication that the size of these effects is small for this specific 
observable.

\subsection{Non-perturbative evaluation of $\Delta_{\rm ls} \aSDO$}
\begin{figure}[t]
	\includegraphics*[width=0.48\linewidth]{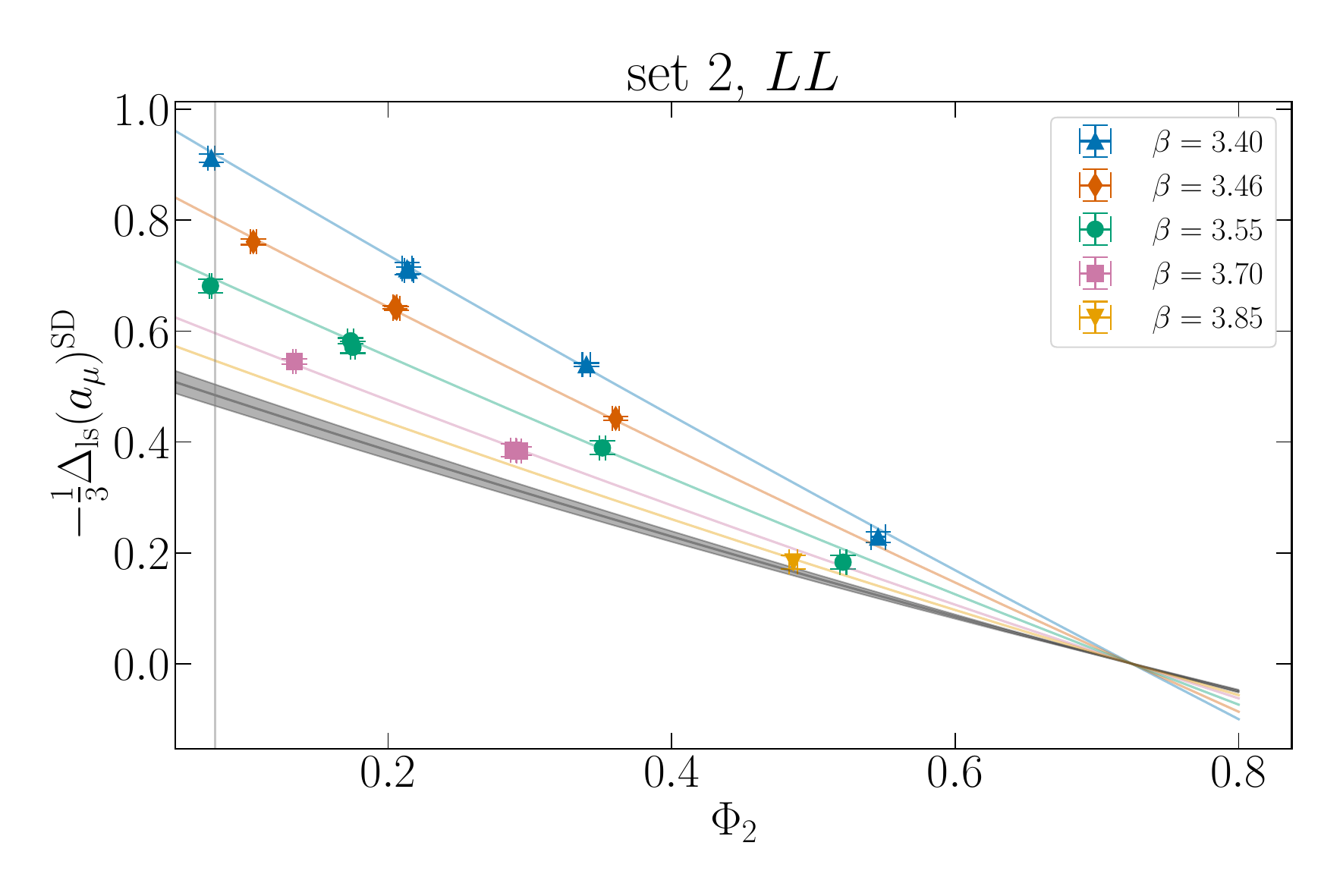}%
	\includegraphics*[width=0.48\linewidth]{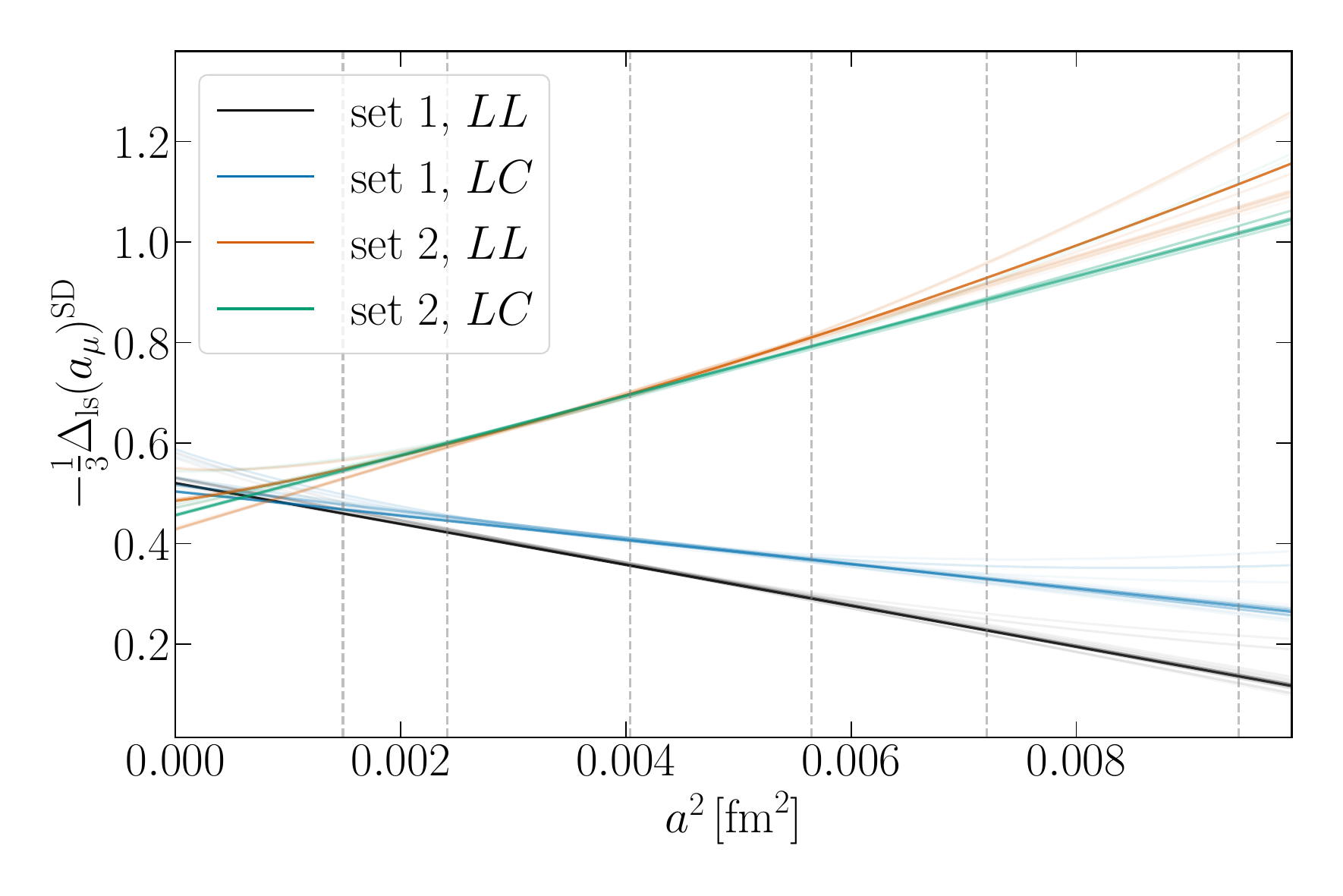}%
	\caption{
		Illustration of fits to $\Delta_{\rm ls} \aSDO$.
		Left: 
		Best fit, according to its model weight, to the data for set 2, 
		$\loc\loc$ with $\chi^2/{\rm dof} = 13.07/13$. 
		The data is corrected for deviation from physical $\Phi_4$.
		The coloured lines show the evaluation of the fit function at finite 
		lattice spacing. The black line, together with the gray band, show the
		dependence on $\Phi_2$ in the continuum limit.
		Right: 
		Projection onto the lattice spacing dependence and physical quark masses for the best fits in the four categories entering the model average. 
	}
	\label{f:aSDId_singlefits}
\end{figure}
To compute the isoscalar contribution to $\aSD$, it suffices to evaluate
$\Delta_{\rm ls} \aSDO$ as given by eq.~(\ref{eq:G88esti}). 
The quark-disconnected loops for light and strange quarks, computed as 
outlined in \cite{Ce:2022eix}, enter here and
may be precisely determined in the short-distance region.
By definition, the splitting $\Delta_{\rm ls} \aSDO$ vanishes at the SU(3)
symmetric point and is expected to depend at leading order on 
$m_{\rm s} - m_{\rm l}$. 
We take this knowledge into account for the chiral and continuum
extrapolation. Defining $\Xdelta = \XK - \frac{3}{2} \Xpi$, we parametrize
\begin{align}\label{e:cc_su3}
	\Delta_{\rm ls}\aSDO(\Xdelta, \XK, \Xa) = \Xdelta \left(
	\gamma_1 + \gamma_2 \Xdelta
	+ \beta_2 \Xa^2 + \beta_3 \Xa^3
	+ \gamma_0 \XK
	\right)\,,
\end{align}
and again build a variety of ansätze by setting coefficients to zero.
The ensembles with SU(3) symmetry are excluded from these fits. 
All cutoff effects are suppressed by $\Xdelta$ close to the SU(3) symmetric
point.

On the left-hand side of figure~\ref{f:aSDId_singlefits} we show the $\Phi_2$
dependence for the data and a typical fit for set 2 and the $\loc\loc$ 
discretization. The constraint $\Delta_{\rm ls}\aSDO = 0$ for $\Phi_4 = \frac{3}{2}\Phi_2$ at
the SU(3) symmetric point is visible by the intersection of the curves which
denote the $\Phi_2$ dependence at finite lattice spacing and in the continuum,
respectively. Towards the physical point, cutoff effects grow since the
suppression is lifted. 
On the right-hand side of the figure, we show the approach to the continuum for
all fits that are contained in the scan over the different models, evaluated at
physical light and strange quark masses. The data from set 1 approach the
continuum value from below whereas the cutoff effects for data set 2 have the
opposite sign.

With the result
\begin{align}
	\label{e:result_aDeltals}
	{\textstyle\frac{1}{3}}\Delta_{\rm ls}\aSDO = -0.495(7)_{\rm stat}(34)_{\rm syst}(4)_{\rm scale}[36]\,,
\end{align}
from an average over the fit models, eq.~(\ref{eq:G88esti}) 
and the result for the isovector contribution from eq.~(\ref{e:result_aSD33}), we arrive at 
\begin{align}
\label{e:result_aSD88}
{\textstyle \frac{1}{3}}\aSDf{8}{8} = 
13.857(14)_{\rm stat}(78)_{\rm syst}(7)_{\rm scale}[81]
\,,
\end{align} 
for the isoscalar contribution. We note that this result has significant 
correlation with the result for $\aSDf{3}{3}$, which is taken into
account when combining both in $\aSD$.

For comparison with other lattice results, we combine the results for the 
isovector and isoscalar contributions in eqs.~(\ref{e:result_aSD33}) and
(\ref{e:result_aSD88}) to compute the strange-connected contribution,
\begin{align}\label{e:result_aSDss}
	{\textstyle\frac{1}{9}}\aSDf{\rm s}{\rm s} = 9.072(10)_{\rm stat}(58)_{\rm syst}(3)_{\rm scale}[60]\,.
\end{align}
The disconnected contribution from light and strange quark loops is found
to be irrelevant with respect to the precision quoted here. A direct evaluation
of the strange-connected contribution via the same strategy as for the 
charm-connected contribution gives a result that is fully compatible 
with eq.~(\ref{e:result_aSDss}).

\subsection{Evaluation of the charm-connected contribution \label{sec:charm_conn}}

\begin{figure}[t]
	\includegraphics*[width=0.48\linewidth]{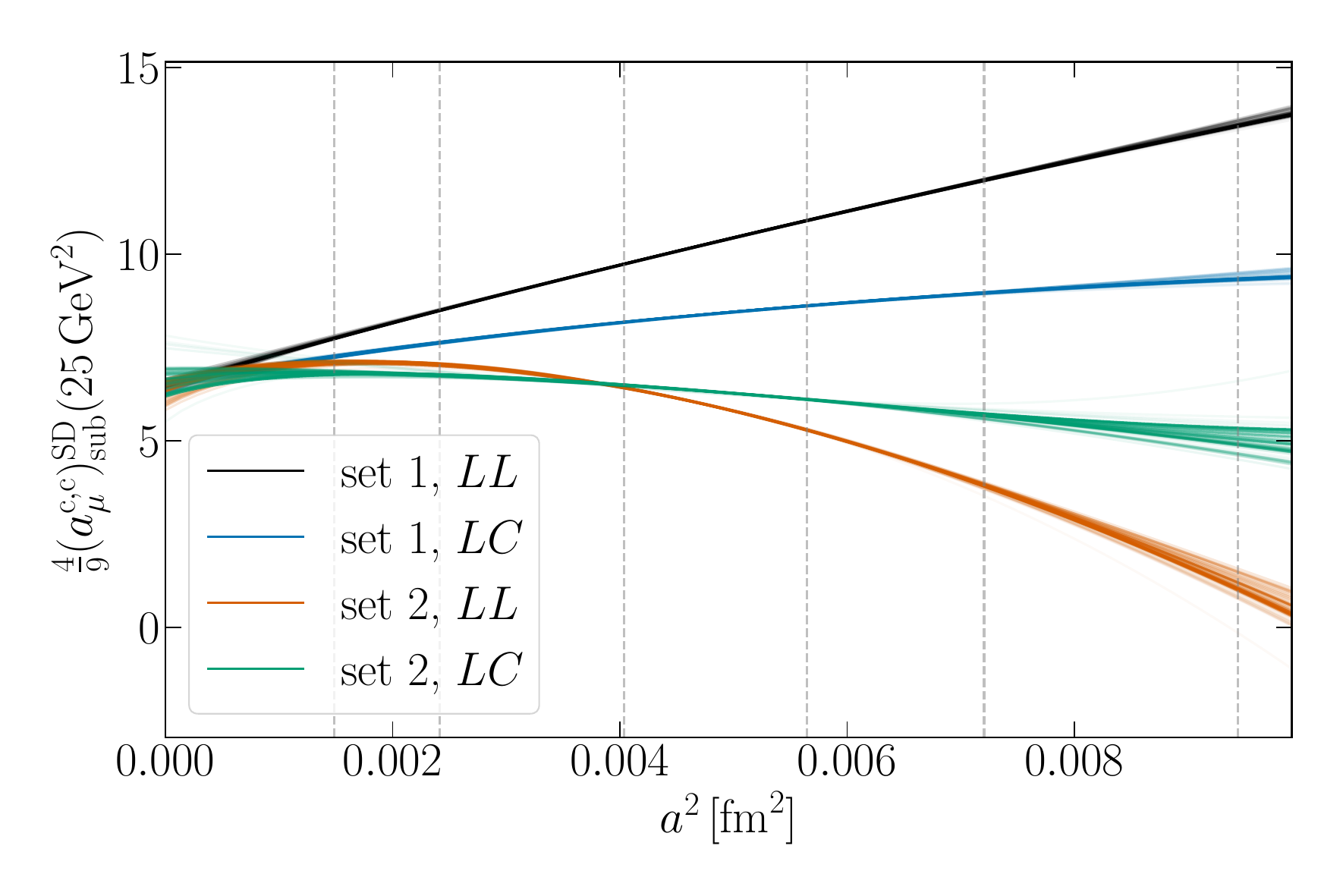}%
	\includegraphics*[width=0.48\linewidth]{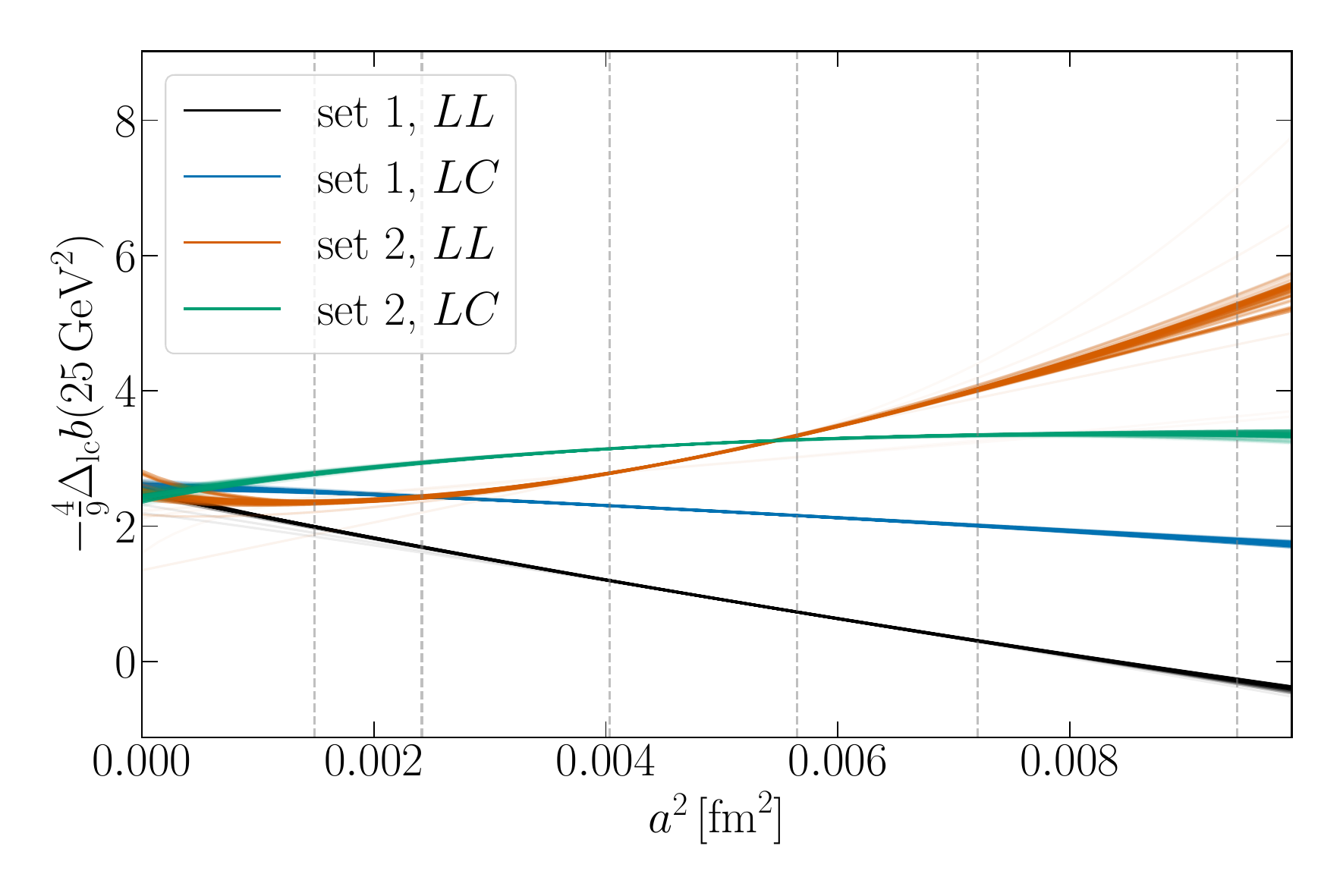}%
	\caption{
		Lattice spacing dependence from a scan over fit models for four 
		data sets. The lattice spacings used in this work are indicated by
		vertical lines.
		Left: fits to $\aSDsubf{\rm c}{\rm c}(36\,{\rm GeV}^2)$.
		Right: fits to $-{\textstyle \frac{4}{9}} \Delta_{\rm lc}\bSD(25\,{\rm GeV}^2)$.
	}
	\label{f:aSDcc_singlefits}
\end{figure}

To compute the charm-connected contribution on the lattice, we start by evaluating
$\aSDsubf{\rm c}{\rm c}(Q^2)$. The tuning of the charm quark
hopping parameter to match the mass of the $D_{\rm s}$ meson of 1968.47\,MeV
and the determination of the renormalization constant for the local charm 
current in a massive renormalization scheme have been described in
\cite{Gerardin:2019rua, Ce:2022kxy}. After the continuum extrapolation, we 
perform a small correction to adapt the tuning to the updated value 
of the scale setting quantity $t_0^{\rm phys}$ from \cite{Strassberger:2021tsu}.

In contrast to the isovector contribution, we find very good agreement of the
continuum extrapolations of the four data sets, despite having significantly
different cutoff effects. We illustrate the cutoff effects based on the scan 
over the fit models for each of the four data sets on the left-hand side  
of Figure~\ref{f:aSDcc_singlefits}. For our final value,
in line with our previous work, we use only the $\loc\cons$ discretization of the 
current, since it exhibits significantly smaller cutoff effects.

The effect of massive quarks in the contribution $\bSDf{\rm c}{\rm c}$ 
may be computed on the lattice without potentially dangerous log-enhanced
cutoff effects. 
The size of this contribution strongly depends on the choice of $Q$ and at large
values of $Q$, the observable is dominated by distances $<0.3\,$fm.
On the right-hand side of  Figure~\ref{f:aSDcc_singlefits} we show the 
approaches to the continuum, based on the scan of the fit models.
We evaluate $\Delta_{\rm lc}\bSD(Q^2)$ for the same values of $Q^2$ as
$\aSDf{3}{3}$ to explore the parameter space and to be able to compare to the
evaluation of the same observable using massive perturbation theory, see table~\ref{t:b_pert}.

\begin{figure}[t]
	\includegraphics*[width=0.48\linewidth]{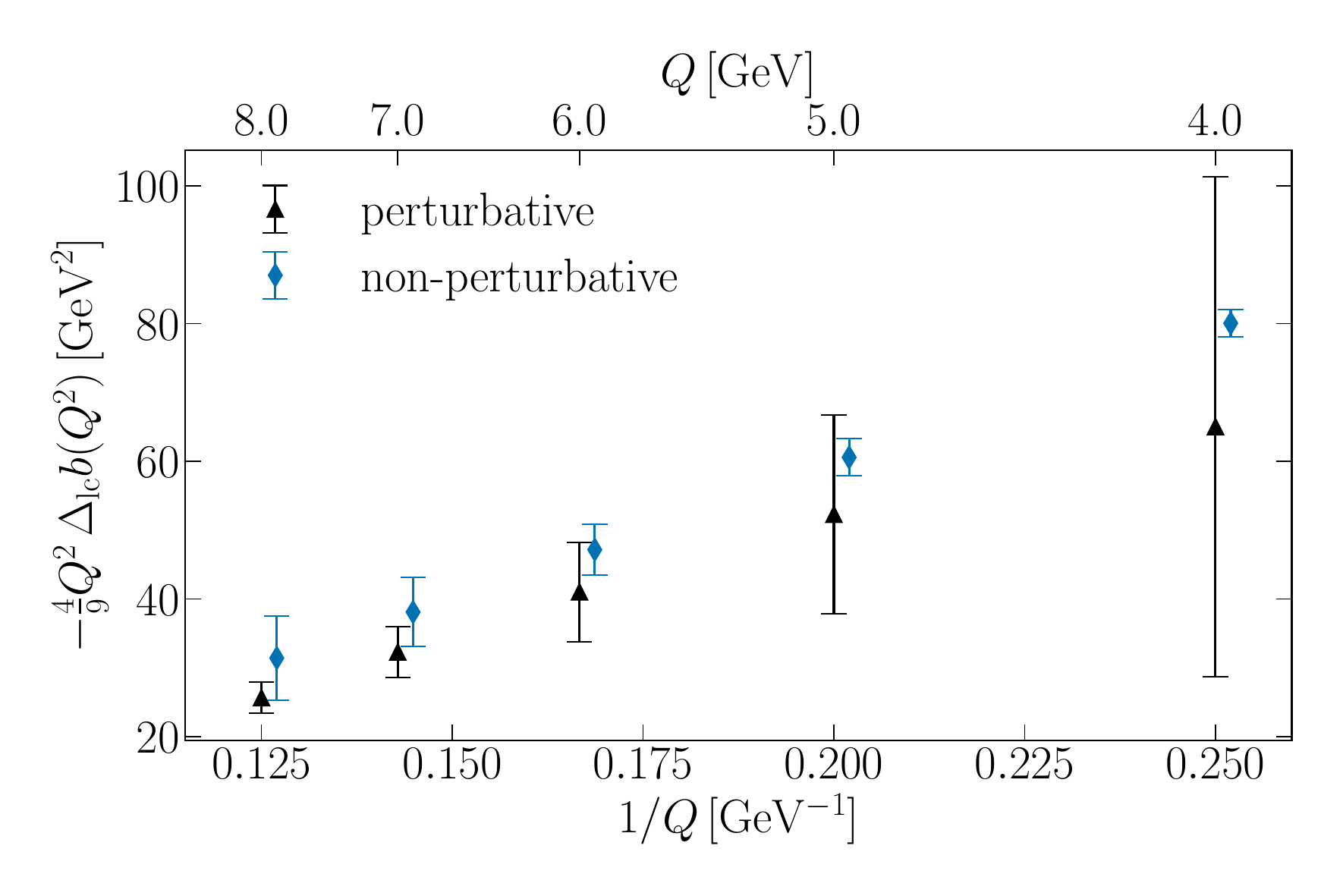} \
	\includegraphics*[width=0.48\linewidth]{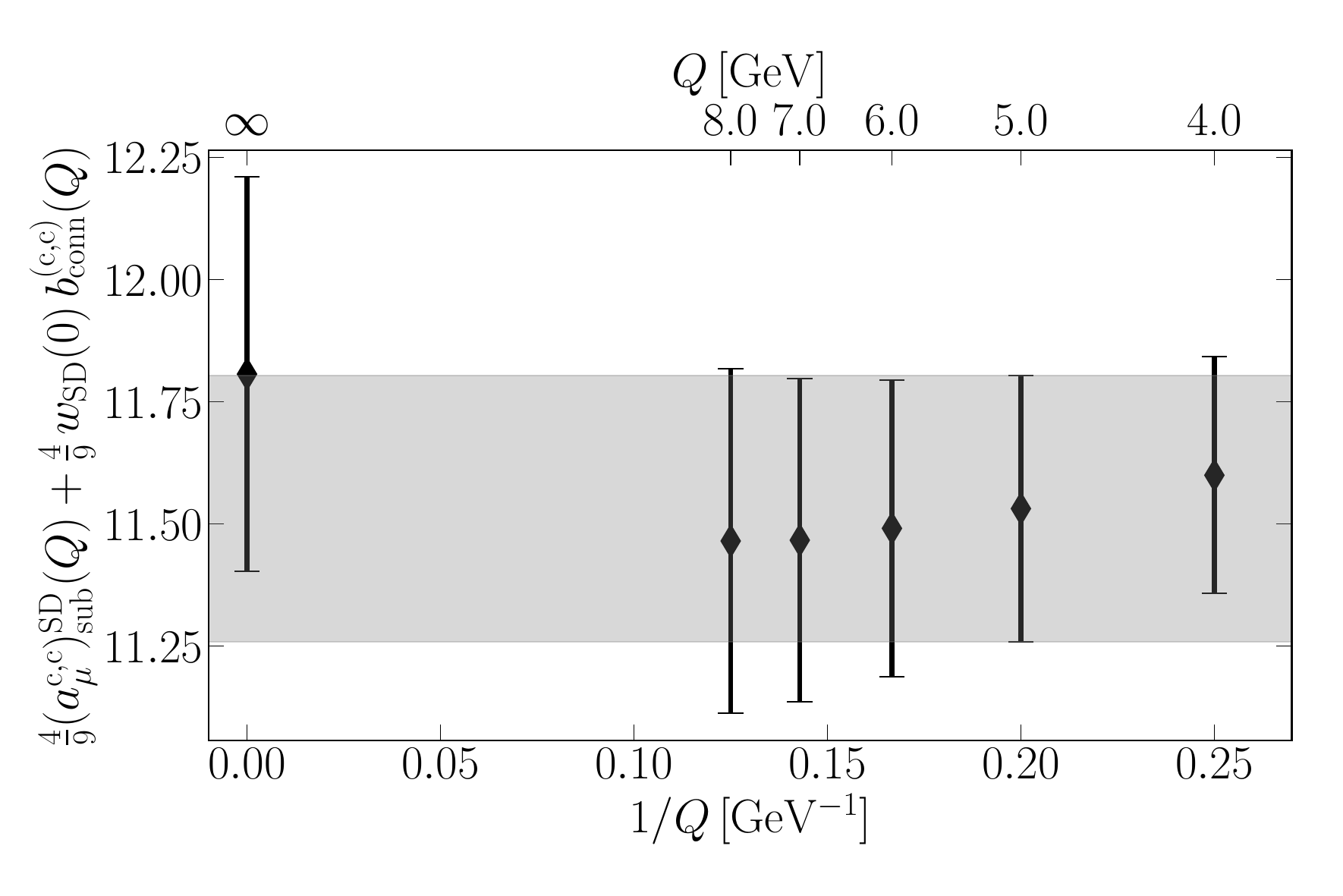} 
	
	\caption{Left: The difference ${-\textstyle\frac{4}{9}}Q^2\,\Delta_{\rm lc}b(Q^2)$ computed non-perturbatively in lattice QCD
	  vs.\ using massive perturbation theory.
	  For visibility, the data is slightly displaced horizontally.
		Right: Stability of the result for $\aSDf{\rm c}{\rm c}$ as a function of the virtuality $Q^2$ chosen in the 
		subtraction term. Choosing $Q=\infty$ corresponds to the entire $\aSDf{\rm c}{\rm c}$ being computed in lattice QCD
		without any use of perturbation theory.  }
	\label{f:b_cc}
\end{figure}

The perturbative evaluation is based on the expansion in $\overline m_c^2/Q^2$ of
the vacuum polarization, computed to O($\alpha_s^2$) in~\cite{Chetyrkin:1997qi}.
We have kept expansion terms up to O($(\overline m_c^2/Q^2)^4$) in the perturbative orders
$\alpha_s^0$ and  $\alpha_s$, while we kept expansion terms up to O($(\overline m_c^2/Q^2)^3$)
in the O($\alpha_s^2$) contribution.
When computing $\bSDf{\rm c}{\rm c}(Q^2)$, we evaluate $\alpha_s$ as well as the
$\overline{\rm MS}$ charm mass
at a scale $\mu^2$ given by the geometric mean of $Q^2$ and $Q^2/4$, using the FLAG'21~\cite{FlavourLatticeAveragingGroupFLAG:2021npn}
value of $\Lambda_{\overline{\rm MS}}$ and starting from the PDG value
$\overline m_c(\mu=\overline m_c) = 1.27\,$GeV~\cite{ParticleDataGroup:2020ssz}.
The number indicated in brackets in the $\bSDf{\rm c}{\rm c}$ column of table \ref{t:b_pert}
represents the contribution of the O($\alpha_s^2$) term
as a way to indicate the uncertainty of the prediction. The quark-mass dependent contribution to $\bSDf{\rm c}{\rm c}(Q^2)$
is compared to lattice data in Fig.\ \ref{f:b_cc}.

On the left-hand side of Figure~\ref{f:b_cc}, we show the comparison of 
the perturbative prediction and the lattice result for 
$\Delta_{\rm lc}\bSD(Q^2)$ in the form proposed in eq.~(\ref{eq:aSDdiff}).
We choose $Q_{\rm ref} = \infty$ where $\Delta_{\rm lc}$ has to vanish.
It is visible that at small $Q$, despite giving similar central values, 
the uncertainty on the perturbative result is significantly larger
than the uncertainty on the lattice evaluation, which is dominated by 
the systematic uncertainty from the continuum extrapolation.
We note that the relative size of this uncertainty grows with increasing $Q$,
whereas its absolute size decreases.

For our final result, we combine the non-perturbative evaluations of 
$\aSDsubf{\rm c}{\rm c}(Q^2)$ and $\Delta_{\rm lc}\bSD(Q^2)$ with the 
perturbative result for $\bSDf{3}{3}(Q^2)$. The combination for several
values of $Q^2$ is shown on the right-hand side of Figure~\ref{f:b_cc}. 
The residual dependence on $Q$ is negligible with respect to the
uncertainties.
For our final result, we again choose $Q=5\,$GeV and with
\begin{align}\label{e:result_aSDsubcc}
	{\textstyle \frac{4}{9}} \aSDsubf{\rm c}{\rm c}(25\,{\rm GeV}^2) 
	&{}= \phantom{-}
	6.81(9)_{\rm stat}(21)_{\rm syst}(7)_{\rm scale}[25]\,,
	\\
	{\textstyle \frac{4}{9}} \Delta_{\rm lc}\bSD(25\,{\rm GeV}^2) 
	&{}= 
	-2.42(4)_{\rm stat}(10)_{\rm syst}(3)_{\rm scale}[11]\,,
\end{align}
we arrive at
\begin{align} \label{e:result_aSDcc}
	{\textstyle \frac{4}{9}} \aSDf{\rm c}{\rm c} = 
	11.53(13)_{\rm stat}(23)_{\rm syst}(11)_{\rm scale}[30]
	\,.
\end{align}

\subsection{The charm-disconnected contributions}
\begin{figure}[t]
	\includegraphics*[width=0.48\linewidth]{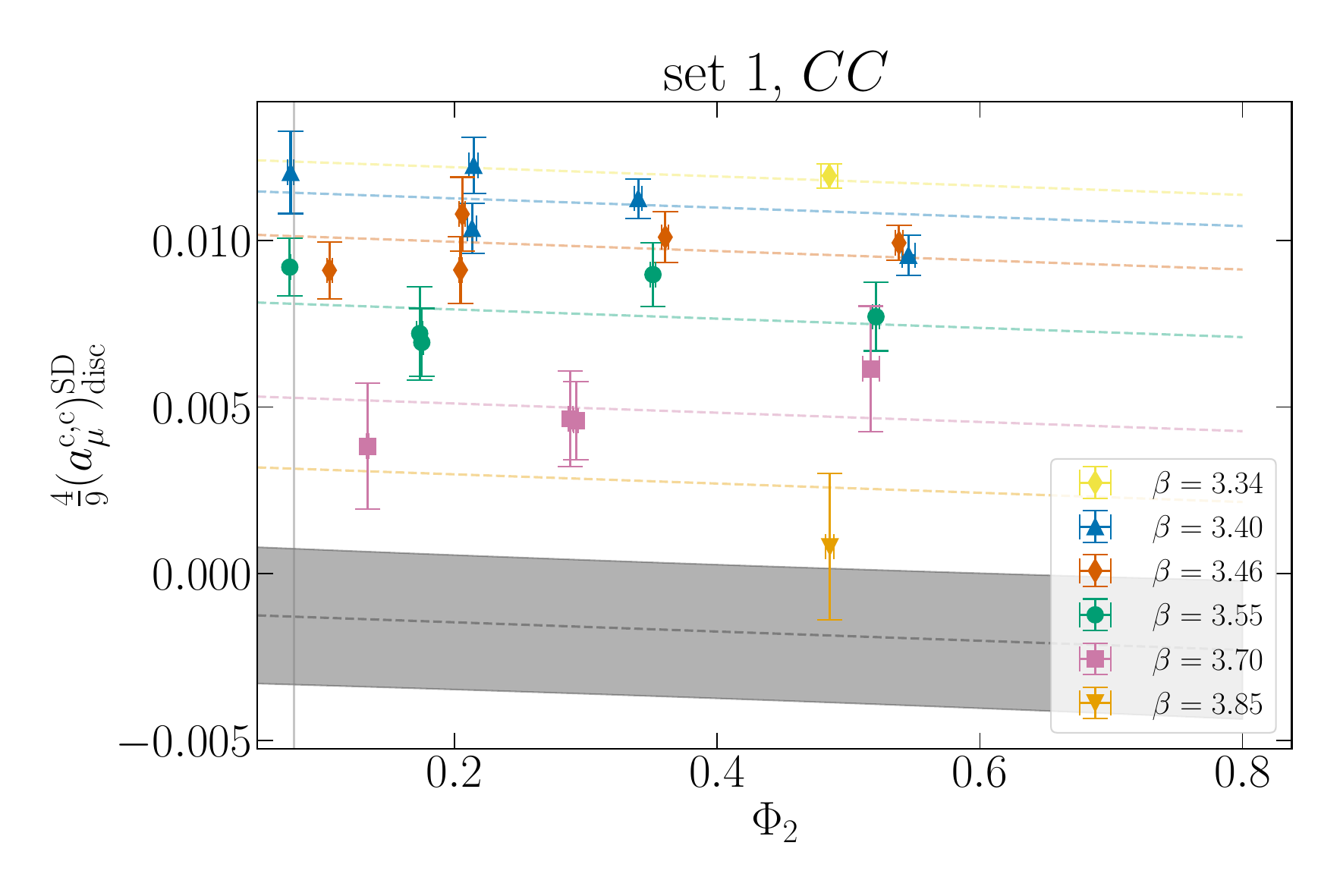}%
	\includegraphics*[width=0.48\linewidth]{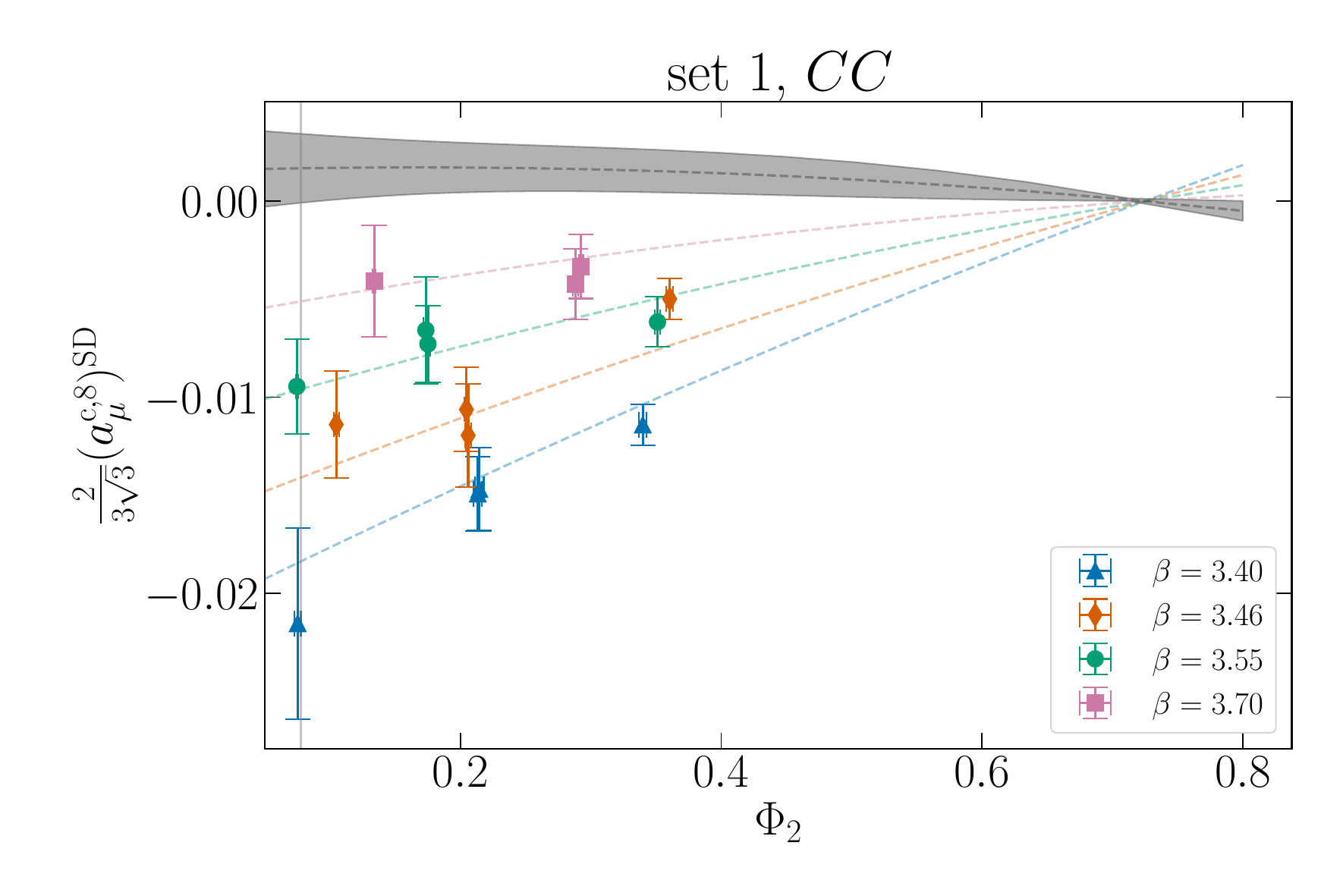}%
	\caption{
		Illustration of typical fits to the charm-disconnected contributions.
		The data is corrected for deviations from $\Phi_4^{\rm phys}$. 
		The coloured lines show evaluations of the fit function at finite lattice spacing and the grey area show the result in the continuum limit.
		Left: Fit to ${\textstyle \frac{4}{9}}\aSDf{\rm c}{\rm c}_{\rm disc}$ with $\chi^2/{\rm dof} = 17.60/16$.
		Right: Fit to ${\textstyle \frac{2}{3\sqrt{3}}}\aSDf{\rm c}{\rm 8}_{\rm disc}$ with $\chi^2/{\rm dof} = 9.20/12$.
	}
	\label{f:aSDdisc}
\end{figure}
The quark-disconnected contributions including charm quarks have not been
included in our previous results for $\ahvp$ \cite{Gerardin:2019rua} and
$(\ahvp)^{\rm ID}$ \cite{Ce:2022kxy} because they are expected to be numerically
very small and mostly contained in the short-distance region. 
In this work, we compute the contributions $\Gf{\rm c}{8}$ and 
$\Gf{\rm c}{\rm c}_{\rm disc}$ in a partially quenched setup, see Appendix C of
\cite{Ce:2022eix} for an explanation of the computational setup. 
The charm quark hopping parameter is set to the same value as in the 
quark-connected case.
To avoid mixing with the singlet current at $\mathrm{O}(a)$ 
in the case of $\Gf{\rm c}{8}$, and anticipating smaller cutoff effects
in the case of a heavy quark (see section \ref{sec:charm_conn}),
we evaluate the correlation functions with 
conserved vector currents at source and sink. 

We find both contributions to be very small and, at finite lattice spacing, 
non-zero only below a distance of about 1\,fm. 
Given the smallness of these contributions,
we perform the analysis without subtractions in the kernel function,
ignoring the possibility of log-enhanced cutoff effects.

The charm-charm contribution is dominated by cutoff effects, already visible 
by comparing different discretizations of the current on a single
ensemble. We find significantly different integrands and even a change in sign
when employing the $\loc\loc$ formulation of the current instead of
$\cons\cons$ or $\loc\cons$. Accordingly, the relative size of the cutoff 
effects is found to be large in the continuum extrapolation.
On the left-hand side of Figure~\ref{f:aSDdisc} we show an exemplary
chiral-continuum extrapolation for this contribution. As can bee seen in the
figure, the sign changes from positive to negative when taking the continuum
limit. We attribute the slight light quark mass dependence to the matching
of the charm quark via the mass of the $D_{\rm s}$ meson and the variation of
the strange quark mass in our set of ensembles.
At the physical point we find after averaging over the fit models the value
\begin{align}\label{e:result_aSDccdisc}
	{\textstyle \frac{4}{9}}\aSDf{\rm c}{\rm c}_{\rm disc} 
	= 
	-0.0010(18)_{\rm stat}(32)_{\rm syst}(1)_{\rm scale}[37]
	\,,
\end{align}
and thus only an upper limit on the magnitude of this contribution.

Since the contribution $\aSDf{c}{8}$ is SU(3) suppressed, we employ fits
according to eq.~(\ref{e:cc_su3}). An example fit is shown on the right-hand
side of Figure~\ref{f:aSDdisc}. Cutoff effects are suppressed close
to the SU(3) symmetric point but significant at physical quark mass. 
The chiral-continuum extrapolated value is compatible with zero.
We obtain
\begin{align}\label{e:result_aSDc8}
	{\textstyle \frac{2}{3\sqrt{3}}}\aSDf{\rm c}{\rm 8}_{\rm disc} 
	= 
	0.0020(17)_{\rm stat}(25)_{\rm syst}(0)_{\rm scale}[31]
	\,,
\end{align}
from averaging over all fit models.

Both contributions are thus negligible compared to $\aSD$ (and our final 
uncertainty). Naturally, this carries over to $\ahvp$. Summing the two
results above and adding the disconnected contribution that enters our 
final result via $\aSDf{8}{8}$, we quote
\begin{align}
	\aSD_{\rm disc} = 0.0013(25)_{\rm stat}(41)_{\rm syst}(5)_{\rm scale}[49]
\end{align}
to allow for a comparison with other works.

\subsection{Isospin-breaking corrections\label{s:ib}}

We apply two complementary methods to determine isospin-breaking
effects in $\aSD$. The first employs massless QCD perturbation theory
to compute the leading QED effects. The second approach is based on
the explicit calculation of isospin-breaking corrections due to
unequal up and down quark masses and electric charges, following the
same method as in our previous publication on the
intermediate-distance window observable~\cite{Ce:2022kxy}.

To compute the leading QED effects at short distances in massless QCD
perturbation theory, we start from the expression for the relative
correction to the Adler function (see~\cite{Kataev:1992dg}, where
the correction is given for the $R$ ratio)
\be\la{eq:QEDcorr}
\frac{D_{\rm QCD,1\,internal\,\gamma}(Q^2)}{D_{\rm QCD}(Q^2)} = \frac{\alpha}{\pi}\cdot
\left(\frac{3}{4} - \frac{\alpha_s}{\pi} + {\rm O}(\alpha_s^2)\right)  \frac{\sum_{f=u,d,s} {\cal Q}_f^4}{\sum_{f=u,d,s} {\cal Q}_f^2}\,.
\ee
Note that in this notation $D_{\rm QCD}(Q^2) = 3 (1+(\alpha_s/\pi) + {\rm O}(\alpha_s^2))\sum_{f=u,d,s}{\cal Q}_f^2 $
is the pure massless QCD expression, while $D_{\rm QCD,1\,internal\,\gamma}(Q^2)$ is the contribution from
massless QCD containing exactly one internal photon line.
The ratio (\ref{eq:QEDcorr}) amounts to $0.51\times 10^{-3}$ for $\alpha_s=0.30$ for the $(u,d,s)$ contribution,
and to $0.57\times 10^{-3}$ if one restricts one's attention to the $(u,d)$ sector.
These are very small corrections indeed, yielding the estimate $[\aSDf{3}{3} + {\txts\frac{1}{3}} \aSDf{8}{8} ]_{\rm 1\,internal\,\gamma} = 0.03$.
For the QED correction to the charm contribution, we use the KS spectral function expressed in terms of the $\overline{\rm MS}$ charm mass.
This means that the latter mass is kept fixed as the electromagnetic interaction is `turned on'.
The free-quark level contribution of the charm is $\frac{4}{9} \aSDf{\rm c}{\rm c}=11.6$ using the  mass of
$\overline m_c(\overline m_c) = 1.27\,$GeV, while the relative QED correction to that is $0.83\times10^{-3}$.  
In absolute terms, this means
\be\la{eq:gam_in_charm}
{\txts\frac{4}{9}}\aSDf{\rm c}{\rm c}({\rm 1\,internal\,\gamma}) = 0.01 .
\ee

For our second approach we have computed $\aSD$ in QCD+QED on a subset
of our ensembles using Monte Carlo reweighting combined with the
leading-order perturbative expansion of QCD+QED about the isosymmetric
theory~\cite{deDivitiis:2013xla, Risch:2021hty, Risch:2019xio,
	Risch:2018ozp, Risch:2017xxe}. The same method was applied in
Ref.~\cite{Ce:2022kxy}. In order to determine the dependence of
isospin-breaking corrections to the short-distance window observable
on the pion mass and the lattice spacing, we have used eight ensembles
(A654, H102, N101, N452, N451, D450, N203 and N200), covering a much
wider range of pion masses ($220-350$\,MeV) and lattice spacings
($0.064-0.097$\,fm) compared to our previous
calculation~\cite{Ce:2022kxy}. As before, we define our hadronic
renormalization scheme in terms of $m_{\pi^0}^2$,
$m_{K^+}^2+m_{K^0}^2-m_{\pi^+}^2$,
$m_{K^+}^2-m_{K^0}^2-m_{\pi^+}^2+m_{\pi^0}^2$ and the fine-structure
constant $\alpha$. Neglecting isospin-breaking effects in the lattice
scale and in the quark sea, as well as quark-disconnected
contributions in the calculation of the relevant correlation
functions, we realize our scheme by matching the values of
$m_{\pi^0}^2$ and $m_{K^+}^2+m_{K^0}^2-m_{\pi^+}^2$ in QCD+QED to
those in the isosymmetric theory and by setting
$m_{K^+}^2-m_{K^0}^2-m_{\pi^+}^2+m_{\pi^0}^2$ to its experimental
value. Further technical details are described in section~VI of
Ref.~\cite{Ce:2022kxy}.

\begin{figure}[t]
	\includegraphics*[width=0.96\textwidth]{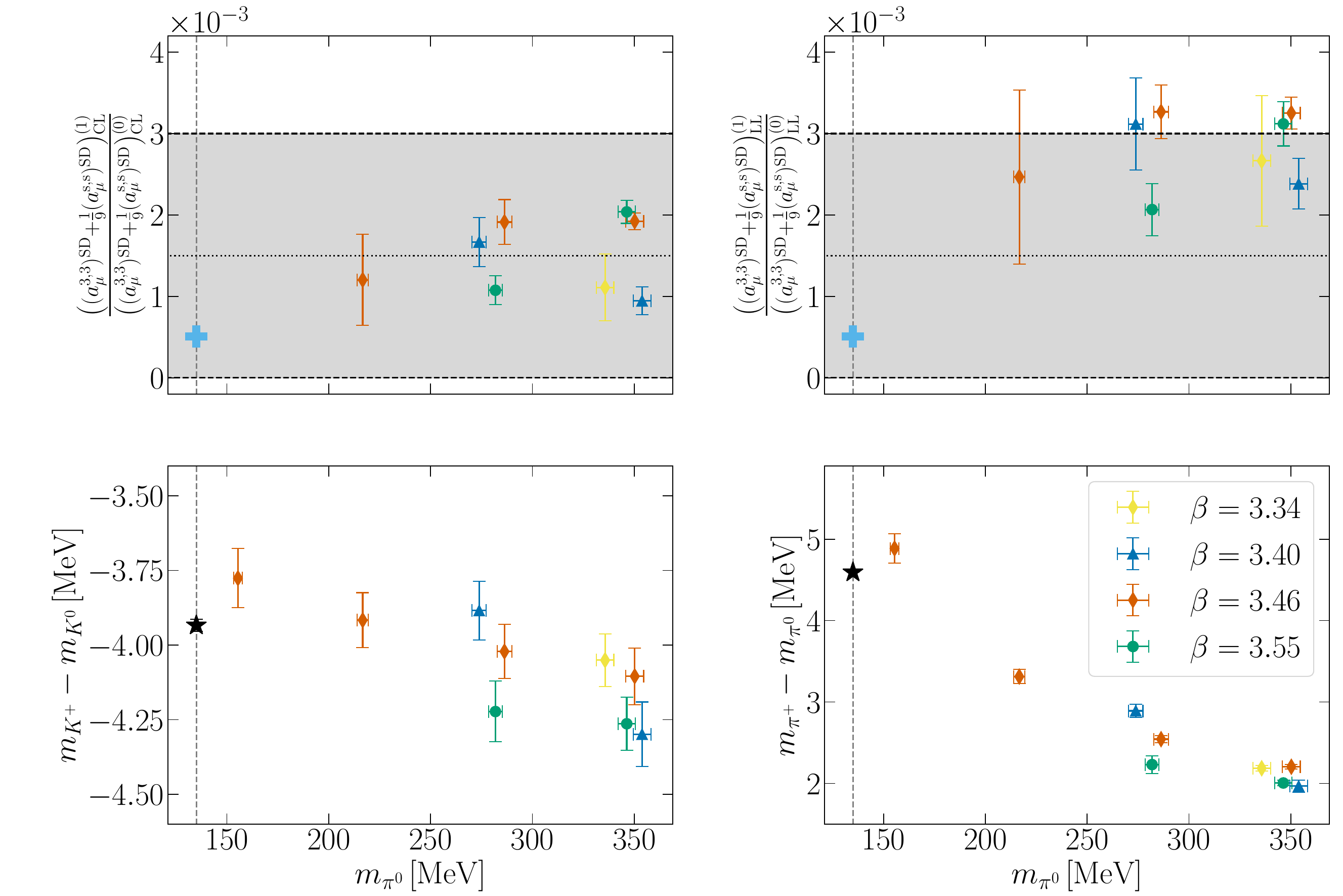}%
	\caption{The relative strong and electromagnetic isospin-breaking
		corrections to the connected light and strange quark contribution
		computed using the conserved-local (upper left panel) and
		local-local (upper right panel) discretizations of the vector
		current. The light blue crosses denote the prediction from massless
		QCD (see the main text). The dashed line and the grey area show 
		our final estimate.
		The two panels at the bottom show the mass splittings of
		charged and neutral kaons and pions, respectively. The black
		stars represent the experimental values.
		}
	\label{f:isospin}
\end{figure}

Our results for the relative size of the isospin-breaking corrections
to the connected light and strange contributions to $\aSD$ are shown
in the upper panel of Figure~\ref{f:isospin} for the conserved-local
(left) and local-local (right) combinations of vector currents. We
make two observations: first, the relative corrections are very small
and typically amount to a few per-mil. Second, there is no clear
dependence of the results on the pion mass and the lattice spacing,
which would allow for a systematic extrapolation to the physical
point. On the other hand, our results for the meson mass splittings
$m_{K^+}-m_{K^0}$ and $m_{\pi^+}-m_{\pi^0}$, shown in the lower panels
of Figure~\ref{f:isospin}, and which include an additional ensemble at
almost physical pion mass (D452), show a consistent trend towards the
corresponding experimental values. We are, therefore, confident that
the typical size of isospin-breaking corrections to $\aSD$ has been
determined correctly. Taking into account the spread in the results
and allowing for a generous error, we represent the data shown in the
upper panels of Figure~\ref{f:isospin} by
$(1.5\pm1.5)\times10^{-3}$. We can then work out the absolute
correction to the connected light and strange quark contribution, by
multiplying the sum of eqs.~(\ref{e:result_aSD33})
and~(\ref{e:result_aSDss}) with 0.0015, which yields \be\la{e:IBls}
[\aSDf{3}{3}+{\textstyle\frac{1}{9}}\aSDf{\rm s}{\rm s}]\times 0.0015
= 0.085\,.  \ee
Adding the QED correction to the charm quark contribution of~0.01 from
eq.~(\ref{eq:gam_in_charm}) and assuming an uncertainty of 100\% for
the latter, we arrive at
\be\la{e:totalIB}
[\aSDf{3}{3}+{\textstyle\frac{1}{9}}\aSDf{\rm s}{\rm s}
+{\txts\frac{4}{9}}\aSDf{\rm c}{\rm c}]_{\rm SIB,\,QED}=0.095(95)\,,
\ee
which we quote as the total isospin-breaking correction to $\aSD$.
This number is larger but of the same order of magnitude compared to
the QED correction to $[\aSDf{3}{3}+{\textstyle\frac{1}{3}}\aSDf{8}{8}
  +{\txts\frac{4}{9}}\aSDf{\rm c}{\rm c}]$ estimated in massless QCD
perturbation theory at leading order and denoted by the light blue
crosses in Fig.~\ref{f:isospin}, which further corroborates our
findings.

\subsection{Further, small contributions \label{s:further_contribs}}

In this section, we present our (mostly perturbative) estimates for
two small contributions that we did not evaluate directly in
lattice QCD.

For the contribution of the $b$ quark, $\aSDf{\rm b}{\rm b}$, the
perturbative spectral function is known exactly to O($\alpha_s$) from
the calculation of K\"all\'en-Sabry (KS)~\cite{Kallen:1955fb},
described for instance in~\cite{Chetyrkin:1996ela}. The O($\alpha_s^2$)
corrections were calculated in~\cite{Chetyrkin:1995ii}.
We have evaluated the KS prediction, improved by substituting the pole quark-mass
appearing in $R(s)$ by its $\overline{\rm MS}$ counterpart and evaluating the latter at the scale $s$.
Using the PDG bottom mass of 4.18\,GeV,
this results in the estimate ${\txts\frac{1}{9}}\aSDf{\rm b}{\rm b} = 0.31$.

We also note the NRQCD based lattice calculation~\cite{Colquhoun:2014ica}, which finds
${\txts\frac{1}{9}}a_\mu^{\rm b,b}=0.271(37)$,
while the phenomenological estimate~\cite{Erler:2021bnl} based on sum rules obtains $0.30(2)$.
We have checked that the bottom contribution to ${\txts\frac{1}{9}}(a_\mu^{\rm b,b} - \aSDf{\rm b}{\rm b})$
is on the order of $0.003$. We thus arrive at our estimate of
\be
{\txts\frac{1}{9}}\aSDf{\rm b}{\rm b} = 0.29(3) .
\ee

As for the effect of the missing charm sea quarks in our lattice calculation,
we estimate its order of magnitude based on perturbation theory, rather than on $D$ meson loops
as we previously did for the intermediate window in~\cite{Ce:2022kxy}, since $\aSD$ involves relatively short distances.
Using the perturbative results of~\cite{Hoang:1994it}, we estimate the effect to be at the level of 
\be\la{eq:pertcharmsea}
[\aSDf{3}{3} + {\txts\frac{1}{3}} \aSDf{8}{8} ]_{\rm pert.\,charm\,sea\,quark\,effect} = 0.02.
\ee
Note that the perturbative estimate of the connected charm contribution is in rather good agreement with
the lattice calculation (see the text above eq.\ (\ref{eq:gam_in_charm})).
Nevertheless, since eq.\ (\ref{eq:pertcharmsea})
represents only the leading perturbative prediction for the effect under scrutiny,
we will conservatively assign an uncertainty of
\be
\Delta [\aSDf{3}{3} + {\txts\frac{1}{3}} \aSDf{8}{8} ]_{\rm charm\,quenching} = 0.10
\ee
to our calculation due to the quenching of the charm quark.
\bigskip

%% file: sec_conclusions.tex
\section{Conclusion}\label{s:concl}
\begin{figure}[t]
	\includegraphics*[height=0.22\textheight]{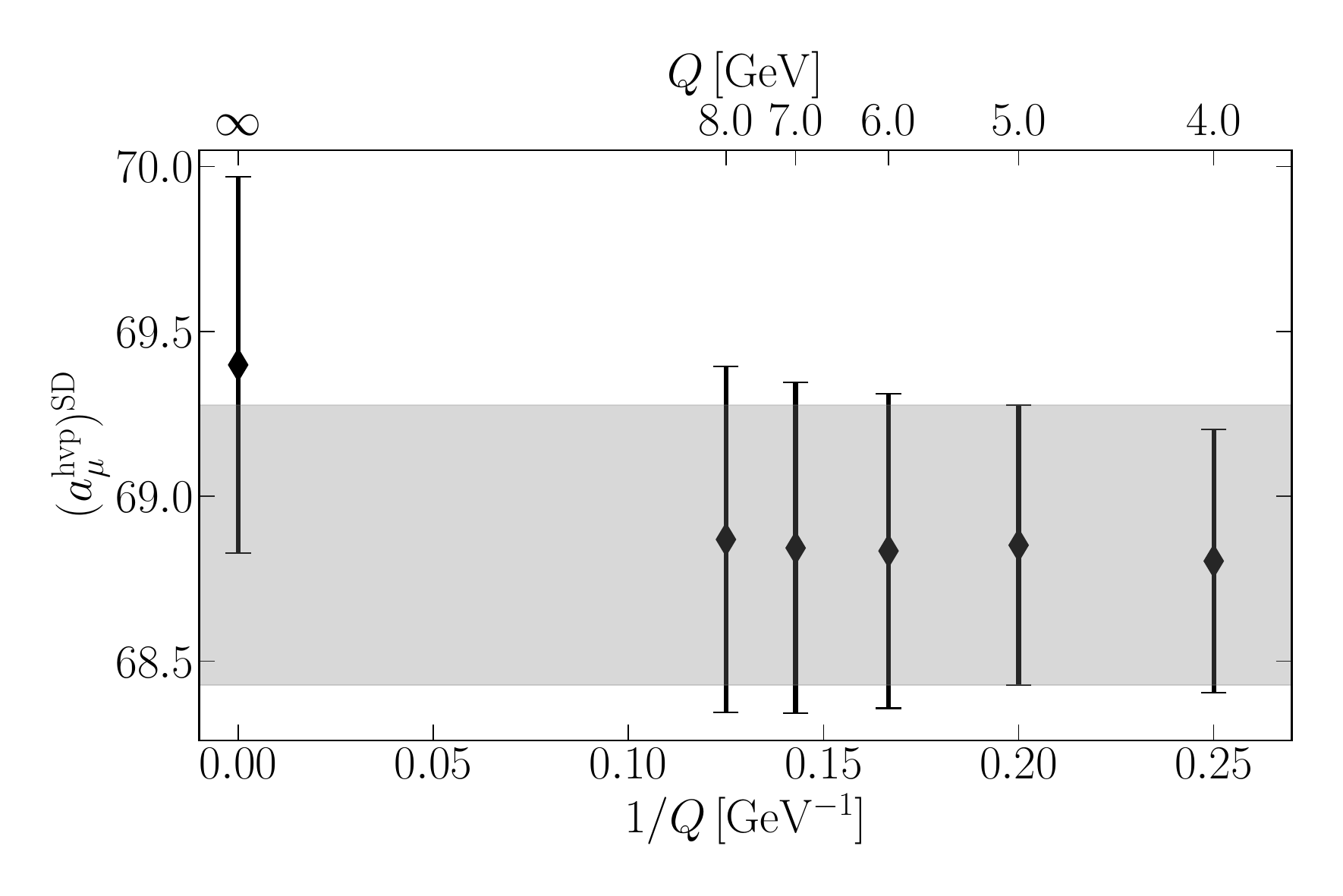}%
	\hspace{.02\linewidth}%
	\includegraphics*[height=0.20\textheight]{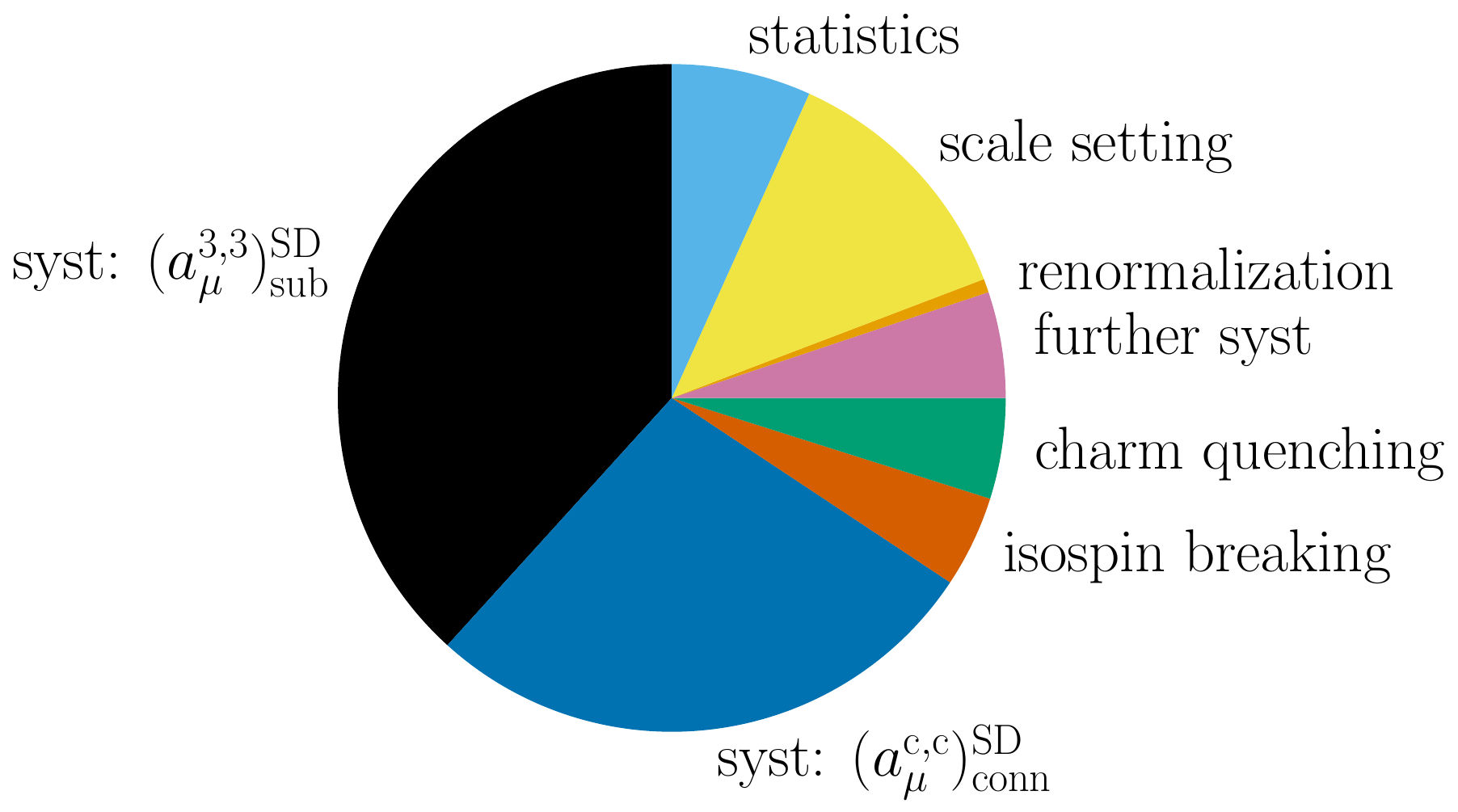}%
	\caption{
	Left: Residual $Q$ dependence of $\aSD$. The grey area shows the final result.
	Right: Contributions to the variance of $\aSD$ as evaluated for
	$Q=5\,$GeV.
	}
	\label{f:aSD_complete}
\end{figure}
Combining the results of sections \ref{s:results},
we arrive at the result
\begin{align}\label{e:result_aSD}
	\aSD = 68.85(14)_{\rm stat}(39)_{\rm syst}(15)_{\rm scale}[45]
	\,,
\end{align}
where all contributions and sources of uncertainty have been taken into account.
As outlined in the main text, we have evaluated the subtracted quantity
of eq.~(\ref{eq:defamuSDsub}) for $Q^2=25\,{\rm GeV}^2$. For this choice,
26\% of the result in eq.~(\ref{e:result_aSD}) are due to massless perturbation theory
applied in the spacelike domain.
On the left-hand side of Figure~\ref{f:aSD_complete} we show results for $\aSD$ for several
choices of the virtuality, the grey band denoting our final estimate.
No residual dependence on the choice for $Q$ can be resolved. At $Q=4\,$GeV
41\% of the result would stem from massless perturbation theory, as given by
Table~\ref{t:b_pert}, whereas the contribution at $Q=8\,$GeV would only be 10\%.
We find a noticeable, but not significant difference of our results using
$\aSDsub$ with respect to a direct evaluation of $\aSD$, as indicated by the
leftmost data point in the figure.

We have shown that, in our data set,
$\aSD$ depends much more strongly on the lattice spacing
than on the quark masses (see Fig.~\ref{f:aSDI1_fits}).
The very fine lattices employed here were thus instrumental in controlling
the continuum limit, even though most of them correspond to heavier-than-physical
pion masses.
The final uncertainty is dominated by systematic uncertainties, 
mostly from the variation of models selected to perform the continuum extrapolation. 
On the right-hand side of Fig.~\ref{f:aSD_complete}, we show the contributions
to the squared uncertainty. The contribution labelled `statistics' corresponds
to the statistical uncertainty of $0.14$ in eq.~(\ref{e:result_aSD}). 
The dependence on the scale setting quantity, and therefore also the resulting
contribution to the final uncertainty, is dominated by the charm quark 
contribution and contains the matching with the mass of the $D_{\rm s}$ meson.

In table~\ref{t:scaledep} we collect the dependence of specific (intermediate)
results with respect to the quantities that define our scheme of isospin
symmetric QCD. The information can be used to adapt the results to a different
scheme, provided that the differences in the input quantities are small.
For example, a shift in the scale setting quantity $\sqrt{t_0^{\rm phys}}$
to the current $N_{\rm f}=2+1+1$ FLAG average with the central value $\sqrt{(t_0)_{N_{\rm f}= 2+1+1}^{\rm phys}} = 0.14292\,$fm
would result in the adapted value $\aSD = 69.00(45)$.

\begin{table}[!t]
	\renewcommand*\arraystretch{1.2}
	\centering
	\begin{tabular}{c|*{5}{c}}
		& \multicolumn{5}{c}{$S$} \\\hline
		\cmidrule(lr){1-6}
		$O$ & $\sqrt{t_0}$ & $m_\pi$ & $m_K$ & $m_{D_{\rm s}}$
		\\\hline
		$\aSDf{3}{3}$ & 
		$ -3.47$ & $ -0.11$ & $ -0.60$ &   --   & \\
		${\textstyle \frac{1}{3}}\aSDf{8}{8}$ & 
		$ -0.70$ & $ +0.04$ & $ -1.55$ &   --   & \\
		${\textstyle \frac{4}{9}} \aSDf{c}{c}$ & 
		$-11.37$ & $ -0.07$ & $ +3.57$ & $-31.05$ & \\
		$\aSD$ & 
		$-15.53$ & $ -0.14$ & $ +1.43$ & $-31.05$ & \\
	\end{tabular}
\caption{Dimensionless scheme dependencies of observable $O$ with respect to 
	the quantity $S$ according to $S \frac{\partial O}{\partial S}$. The four quantities $S$ define the scheme of isoQCD in this work. Their central values are
	$\sqrt{t_0} = 0.1443\,$fm, 
	$m_\pi = 134.9768\,$MeV,
	$m_K = 495.011\,$MeV,
	$m_{D_{\rm s}} = 1968.47\,$MeV.
	\label{t:scaledep}
}
\end{table}

\begin{figure}
\includegraphics*[width=0.75\textwidth]{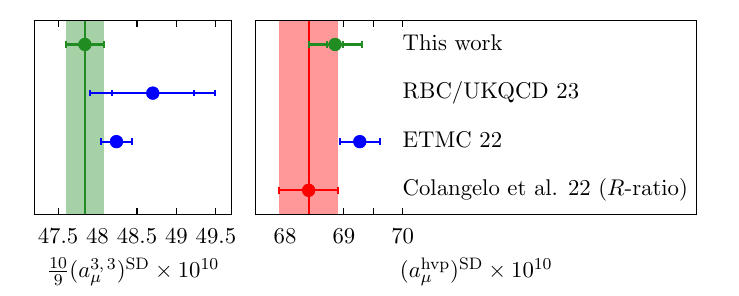}
\vspace{-0.2cm}
\caption{Comparison of our results for the short-distance window
  observable $\aSD$ (right panel) and the light-quark connected
  contribution (left panel) to other recent lattice calculations by
  ETM \cite{ExtendedTwistedMass:2022jpw} and RBC/UKQCD
  \cite{RBC:2023pvn}. Our results are denoted by green circles and
  band, while the data-driven estimate of Colangelo et
  al. \cite{Colangelo:2022vok} is shown as the red circle and band.}
  \label{f:comparison}
\end{figure}

A compilation of recent results for the short-distance window
observable
\cite{Colangelo:2022vok,ExtendedTwistedMass:2022jpw,RBC:2023pvn},
including the estimates from this work, is shown in
Figure~\ref{f:comparison}. Our result (\ref{e:result_aSD}) is in good
agreement with the dispersive estimate from~\cite{Colangelo:2022vok}, $\aSD=68.4(5)$.
Our result is also in good agreement with the lattice result from the ETM
collaboration~\cite{ExtendedTwistedMass:2022jpw}, $\aSD=69.27(34)$.
In the introduction, a schematic scenario was discussed in which the
$R$ ratio corresponding to lattice calculations is enhanced by six
percent relative to the $e^+e^-$ data entering the determination
of~\cite{Colangelo:2022vok} in the interval $600\,{\rm
  MeV}<\sqrt{s}<900\,{\rm MeV}$. Such an enhancement would increase
$\aSD$ by about one unit.  Both the ETM and our result are also
consistent with this scenario, which thus remains a working hypothesis
worth challenging.

Our result for the light-quark connected contribution,
$\frac{10}{9}\aSDf{3}{3}=47.84(24)$ is consistent with the ETM
result~\cite{ExtendedTwistedMass:2022jpw} $48.24(20)$ and the
RBC/UKQCD result~\cite{RBC:2023pvn} $48.70(52)(59)$ at their favoured
isospin-symmetric point.  The dependence of $\aSD$ on the precise
choice of this point is currently smaller than the statistical errors. 

Finally, the subtraction strategy for the short-distance effects
adopted here can be reused for the full $\ahvp$ (the analogue of
eq.\ (\ref{eq:amuSDfinalrep}) being $\ahvp = (\ahvp)_{\rm sub}(Q^2) +
\bSDf{\gamma}{\gamma}(Q^2)$), as well as for other observables related
to the vacuum polarization, such as the running of the electromagnetic
coupling. The appeal of this subtraction is that it removes the
leading, O($x_0^4$) term from the kernel, while at the same being safe
to compute in perturbation theory, since it only involves spacelike photon
virtualities on the order of $Q^2$.

%% file: app_tables.tex
\section{Tables}
\input{./tables/tab_ps.tex}
\input{./tables/tab_fvc.tex}
\input{./tables/tab_5gev.tex}
\input{./tables/tab_ls.tex}
\input{./tables/tab_lc.tex}
\input{./tables/tab_disc.tex}

%% file: tables/tab_ps.tex
\begin{table}[!ph]
	\caption{Pseudoscalar masses in lattice units, including finite-size corrections. Value of the gluonic observable $t_0/a^2$ and the two dimensionless variables $\phi_2$ and $\phi_4$ used in the extrapolation to the physical point.}
	\vskip 0.1in
	\renewcommand{\arraystretch}{1.1}  
	\begin{tabular}{l|@{\hskip 01em}c@{\hskip 01em}c|@{\hskip 01em}c@{\hskip 01em}c@{\hskip 01em}c@{\hskip 01em}c
		}
		\hline
		id      &        $a m_{\pi}$    &       $a m_{K}$               &       $t_0/a^2$      &       $\phi_2$ &       $\phi_4$ \\
		\hline
		A653 & 0.21183(105) & 0.21183(105) & 2.173(07)  & 0.7801(58) & 1.1701(88)  \\
		A654 & 0.16632(131) & 0.22727(112) & 2.194(10)  & 0.4854(61) & 1.1491(84)  \\
		\hline
		H101 & 0.18250(71)  & 0.18250(71)  & 2.847(06)  & 0.7586(45) & 1.1379(68)  \\
		H102 & 0.15383(80)  & 0.19135(71)  & 2.882(12)  & 0.5457(50) & 1.1172(75)  \\
		H105 & 0.12154(115) & 0.20223(85)  & 2.886(09)  & 0.3411(56) & 1.1149(83)  \\
		N101 & 0.12120(56)  & 0.20146(35)  & 2.892(03)  & 0.3399(29) & 1.1090(39)  \\
		C101 & 0.09570(78)  & 0.20584(44)  & 2.913(05)  & 0.2134(33) & 1.0940(48)  \\
		C102 & 0.09671(78)  & 0.21761(47)  & 2.868(05)  & 0.2146(33) & 1.1939(45)  \\
		D150 & 0.05654(94)  & 0.20835(35)  & 2.944(04)  & 0.0753(25) & 1.0600(34)  \\
		\hline
		B450 & 0.16081(50)  & 0.16081(50)  & 3.663(13)  & 0.7578(35) & 1.1367(53)  \\
		S400 & 0.13503(46)  & 0.17022(41)  & 3.692(08)  & 0.5385(31) & 1.1250(46)  \\
		N451 & 0.11064(45)  & 0.17822(26)  & 3.682(07)  & 0.3606(25) & 1.1158(31)  \\
		D450 & 0.08346(51)  & 0.18393(26)  & 3.697(06)  & 0.2060(23) & 1.1036(32)  \\
		D451 & 0.08359(30)  & 0.19402(14)  & 3.662(04)  & 0.2047(14) & 1.2051(16)  \\
		D452 & 0.05932(59)  & 0.18645(18)  & 3.727(04)  & 0.1049(21) & 1.0888(22)  \\
		\hline
		H200 & 0.13625(64)  & 0.13625(64)  & 5.151(33)  & 0.7649(81) & 1.1474(121) \\
		N202 & 0.13445(42)  & 0.13445(42)  & 5.158(19)  & 0.7459(37) & 1.1188(56)  \\
		N203 & 0.11249(27)  & 0.14395(23)  & 5.146(08)  & 0.5209(24) & 1.1136(36)  \\
		N200 & 0.09221(29)  & 0.15065(24)  & 5.163(07)  & 0.3512(20) & 1.1130(33)  \\
		D251 & 0.09203(16)  & 0.15041(12)  & 5.164(05)  & 0.3499(11) & 1.1096(16)  \\
		D200 & 0.06502(28)  & 0.15630(17)  & 5.179(06)  & 0.1752(15) & 1.0998(23)  \\
		D201 & 0.06499(43)  & 0.16309(24)  & 5.136(08)  & 0.1736(22) & 1.1797(33)  \\
		E250 & 0.04236(23)  & 0.15936(08)  & 5.202(04)  & 0.0747(07) & 1.0943(11)  \\
		\hline
		N300 & 0.10574(30)  & 0.10574(30)  & 8.560(32)  & 0.7657(46) & 1.1486(69)  \\
		N302 & 0.08707(54)  & 0.11363(46)  & 8.526(25)  & 0.5171(63) & 1.1392(102) \\
		J303 & 0.06467(22)  & 0.11963(19)  & 8.618(14)  & 0.2883(19) & 1.1309(36)  \\
		J304 & 0.06559(20)  & 0.13187(17)  & 8.500(14)  & 0.2925(16) & 1.3288(33)  \\
		E300 & 0.04407(15)  & 0.12386(12)  & 8.619(06)  & 0.1339(09) & 1.1248(22)  \\
		\hline
		J500 & 0.08157(17)  & 0.08157(17)  & 13.964(31) & 0.7432(32) & 1.1148(48)  \\
		J501 & 0.06590(23)  & 0.08796(24)  & 13.984(49) & 0.4859(31) & 1.1084(55)  \\
		\hline
	\end{tabular} 
	\label{tab:ps}
\end{table}

%% file: tables/tab_fvc.tex
 \begin{table}[!p]
 	\caption{Finite-size effects for $\aSD$ in the isovector channel, in units of $10^{-10}$. The correction is obtained using the Hansen-Patella method and split according to eq.~(\ref{eq:amuSDfinalrep}). The first part enters our calculation in the non-perturbative determination of $\aSDsubf{3}{3}(25\,{\rm GeV}^2)$, the second one in the determination of $ \Delta_{\rm lc}\bSD(25\,{\rm GeV}^2)$.
 	The columns contain the correction due to pions and kaons wrapping around the torus. On ensembles with SU(3) symmetry, the correction is contained in the correction due to pions. The column labelled `total' gives the sum of the two contributions.
 	When including the finite-volume corrections in our analysis, we assign
 	a flat 25\% relative uncertainty to the correction.
	 }
 	\vskip 0.1in
 	\renewcommand{\arraystretch}{1.1}  
 	\begin{tabular}{l|@{\hskip 01em}c@{\hskip 01em}c@{\hskip 01em}c|@{\hskip 01em}c@{\hskip 01em}c@{\hskip 01em}c@{\hskip 01em}c
 		}
 		\hline
 		& \multicolumn{3}{c|@{\hskip 1em}}{Correction for $\aSDsubf{3}{3}(25\,{\rm GeV}^2)$} 
 		& \multicolumn{3}{c@{\hskip 1em}}{Correction for $w_{\rm SD}(0)\,\bSDf{3}{3}(25\,{\rm GeV}^2)$}
 		\\
 		\hline
 		id      &        
 		pion    &       kaon       &  total    &
 		pion       &    kaon       &  total    \\
 		\hline
  A653 & 0.1035(24)  &     -     & 0.1035(24)  &  0.00580(13)  &     -     & 0.00580(13)  \\
  A654 & 0.1608(55)  & 0.0262(7) & 0.1871(62)  &  0.00910(33)  & 0.0015(0) & 0.01057(35)  \\
  \hline
  H101 & 0.0415(07)  &     -     & 0.0415(07)  &  0.00234(04)  &     -     & 0.00234(04)  \\
  H102 & 0.0576(07)  & 0.0112(1) & 0.0688(08)  &  0.00328(04)  & 0.0006(0) & 0.00391(04)  \\
  H105 & 0.1271(53)  & 0.0085(2) & 0.1356(55)  &  0.00732(31)  & 0.0005(0) & 0.00780(32)  \\
  N101 & 0.0116(03)  & 0.0003(0) & 0.0118(03)  &  0.00069(02)  & 0.0000(0) & 0.00071(02)  \\
  C101 & 0.0310(11)  & 0.0002(0) & 0.0312(11)  &  0.00189(07)  & 0.0000(0) & 0.00190(07)  \\
  C102 & 0.0293(09)  & 0.0001(0) & 0.0294(09)  &  0.00179(06)  & 0.0000(0) & 0.00180(06)  \\
  D150 & 0.0337(15)  & 0.0000(0) & 0.0337(15)  &  0.00220(10)  & 0.0000(0) & 0.00220(10)  \\
  \hline
  B450 & 0.0934(13)  &     -     & 0.0934(13)  &  0.00524(08)  &     -     & 0.00524(08)  \\
  S400 & 0.1188(13)  & 0.0249(2) & 0.1437(16)  &  0.00672(08)  & 0.0014(0) & 0.00811(09)  \\
  N451 & 0.0230(06)  & 0.0008(0) & 0.0239(06)  &  0.00136(03)  & 0.0000(0) & 0.00141(03)  \\
  D450 & 0.0118(04)  & 0.0000(0) & 0.0118(04)  &  0.00074(02)  & 0.0000(0) & 0.00074(02)  \\
  D451 & 0.0116(02)  & 0.0000(0) & 0.0116(02)  &  0.00073(01)  & 0.0000(0) & 0.00073(01)  \\
  D452 & 0.0399(15)  & 0.0000(0) & 0.0399(15)  &  0.00254(10)  & 0.0000(0) & 0.00254(10)  \\
  \hline
  H200 & 0.2425(39)  &     -     & 0.2425(39)  &  0.01356(21)  &     -     & 0.01356(21)  \\
  N202 & 0.0202(03)  &     -    & 0.0202(03)  &  0.00115(02)  &     -     & 0.00115(02)  \\
  N203 & 0.0309(04)  & 0.0046(1) & 0.0356(04)  &  0.00178(03)  & 0.0003(0) & 0.00204(03)  \\
  N200 & 0.0662(08)  & 0.0036(0) & 0.0698(08)  &  0.00385(04)  & 0.0002(0) & 0.00405(04)  \\
  D251 & 0.01122(09) & 0.0003(0) & 0.01148(09) &  0.000674(05) & 0.0000(0) & 0.000689(05) \\
  D200 & 0.0431(06)  & 0.0002(0) & 0.0432(06)  &  0.00264(04)  & 0.0000(0) & 0.00265(04)  \\
  D201 & 0.0427(10)  & 0.0001(0) & 0.0428(10)  &  0.00262(06)  & 0.0000(0) & 0.00263(06)  \\
  E250 & 0.0183(03)  & 0.0000(0) & 0.0183(03)  &  0.00123(02)  & 0.0000(0) & 0.00123(02)  \\
  \hline
  N300 & 0.1027(09)  &     -     & 0.1027(09)  &  0.00576(05)  &     -     & 0.00576(05)  \\
  N302 & 0.1358(36)  & 0.0254(6) & 0.1612(41)  &  0.00769(22)  & 0.0014(0) & 0.00911(25)  \\
  J303 & 0.0768(09)  & 0.0024(0) & 0.0792(09)  &  0.00450(04)  & 0.0001(0) & 0.00463(04)  \\
  J304 & 0.0726(08)  & 0.0012(0) & 0.0738(08)  &  0.00426(05)  & 0.0001(0) & 0.00433(05)  \\
  E300 & 0.0293(03)  & 0.0000(0) & 0.0293(03)  &  0.00185(02)  & 0.0000(0) & 0.00186(02)  \\
  \hline
  J500 & 0.0836(11)  &     -     & 0.0836(11)  &  0.00470(06)  &     -     & 0.00470(06)  \\
  J501 & 0.1210(24)  & 0.0204(7) & 0.1414(30)  &  0.00687(13)  & 0.0011(0) & 0.00801(16)  \\
		\hline
\end{tabular} 
\label{tab:fvc}
\end{table}

%% file: tables/tab_5gev.tex
\begin{table}[!p]
\caption{Values of the subtracted isovector and charm-connected contributions
	 with $Q^2=25\,{\rm GeV}^2$
in units of $10^{-10}$, for the local-local (${\scriptstyle\rm LL}$) and for the local-conserved (${\scriptstyle\rm CL}$) discretizations of the correlation function, as described in the main text.  
The finite-size correction has been applied to the isovector contribution. }
\vskip 0.1in
\renewcommand{\arraystretch}{1.1}  
\begin{tabular}{l|@{\hskip 0.5em}c@{\hskip 0.5em}c@{\hskip 0.5em}|@{\hskip 0.5em}c@{\hskip 0.5em}c@{\hskip 0.5em}|@{\hskip 0.5em}c@{\hskip 0.5em}c@{\hskip 0.5em}|@{\hskip 0.5em}c@{\hskip 0.5em}c}
\hline
        & \multicolumn{2}{c|@{\hskip 0.5em}}{$\aSDsubf{3}{3}$ - Set 1} 
        & \multicolumn{2}{c|@{\hskip 0.5em}}{$\aSDsubf{3}{3}$ - Set 2}
        & \multicolumn{2}{c|@{\hskip 0.5em}}{${\textstyle \frac{4}{9}}\aSDsubf{\rm c}{\rm c}$ - Set 1} 
        & \multicolumn{2}{c}{${\textstyle \frac{4}{9}}\aSDsubf{\rm c}{\rm c}$ - Set 2} \\
\hline
id      &        ${\scriptstyle(LL)}$   &       ${\scriptstyle(CL)}$     &
                 ${\scriptstyle(LL)}$   &       ${\scriptstyle(CL)}$     &      
                 ${\scriptstyle(LL)}$   &       ${\scriptstyle(CL)}$    &       
                 ${\scriptstyle(LL)}$   &       ${\scriptstyle(CL)}$    \\
\hline
 A653 & 39.138(75) & 40.010(45)  & 28.547(47)      & 34.913(31)       & 12.892(38) & 8.860(20)   & 1.240(25)       & 4.875(14)        \\
 A654 & 39.171(82) & 40.102(56)  & 29.198(60)      & 35.484(44)       & 13.039(62) & 9.061(36)   & 1.361(29)       & 5.005(23)        \\
 \hline
 H101 & 38.650(41) & 39.252(30)  & 31.316(28)      & 35.656(25)       & 11.406(43) & 8.455(27)   & 3.683(29)       & 5.405(20)        \\
 H102 & 38.668(40) & 39.293(29)  & 31.712(47)      & 35.985(36)       & 11.503(56) & 8.556(37)   & 3.754(34)       & 5.479(27)        \\
 H105 & 38.695(63) & 39.357(58)  & 32.018(58)      & 36.279(56)       & --         & --          & --              & --               \\
 N101 & 38.823(34) & 39.495(22)  & 32.144(17)      & 36.412(22)       & 11.687(41) & 8.736(25)   & 3.881(29)       & 5.611(19)        \\
 C101 & 38.782(38) & 39.461(26)  & 32.366(24)      & 36.588(20)       & 11.740(41) & 8.797(28)   & 3.924(30)       & 5.654(21)        \\
 C102 & 38.834(36) & 39.526(24)  & 32.314(18)      & 36.605(16)       & --         & --          & --              & --               \\
 D150 & 38.738(35) & 39.427(22)  & 32.660(18)      & 36.832(14)       & --         & --          & --              & --               \\
 \hline
 B450 & 37.916(31) & 38.351(30)  & 32.874(31)      & 35.804(29)       & 10.313(36) & 8.058(24)   & 5.018(30)       & 5.729(20)        \\
 S400 & 37.993(38) & 38.445(37)  & 33.210(41)      & 36.106(38)       & 10.245(45) & 8.126(30)   & 5.024(34)       & 5.789(25)        \\
 N451 & 38.202(15) & 38.678(13)  & 33.582(16)      & 36.485(13)       & --         & --          & --              & --               \\
 D450 & 38.232(11) & 38.722(08)  & 33.817(12)      & 36.703(08)       & 10.684(32) & 8.478(21)   & 5.315(29)       & 6.062(18)        \\
 D451 & 38.265(12) & 38.762(08)  & 33.806(12)      & 36.723(09)       & --         & --          & --              & --               \\
 D452 & 38.218(15) & 38.713(13)  & 33.995(15)      & 36.853(13)       & 10.790(31) & 8.558(21)   & 5.384(30)       & 6.124(18)        \\
 \hline
 H200 & 36.992(82) & 37.229(82)  & 34.122(81)      & 35.725(80)       & 9.133(49)  & 7.633(39)   & 6.045(34)       & 6.075(32)        \\
 N202 & 37.169(32) & 37.409(32)  & 34.274(34)      & 35.878(33)       & 9.159(41)  & 7.657(31)   & 6.064(33)       & 6.095(27)        \\
 N203 & 37.303(24) & 37.561(22)  & 34.531(26)      & 36.140(23)       & 9.279(33)  & 7.762(25)   & 6.168(28)       & 6.188(22)        \\
 N200 & 37.383(29) & 37.647(28)  & 34.764(30)      & 36.361(28)       & 9.445(25)  & 7.929(18)   & 6.306(22)       & 6.331(16)        \\
 D251 & 37.437(12) & 37.703(10)  & 34.797(14)      & 36.395(10)       & --         & --          & --              & --               \\
 D200 & 37.482(23) & 37.757(22)  & 34.992(24)      & 36.582(22)       & 9.605(34)  & 8.083(28)   & 6.434(29)       & 6.461(24)        \\
 D201 & 37.483(19) & 37.764(18)  & 34.968(21)      & 36.577(19)       & --         & --          & --              & --               \\
 E250 & 37.493(10) & 37.773(08)  & 35.122(13)      & 36.703(09)       & 9.671(25)  & 8.162(19)   & 6.493(24)       & 6.530(18)        \\
 \hline
 N300 & 36.064(36) & 36.124(41)  & 34.866(35)      & 35.432(40)       & 7.916(41)  & 7.074(35)   & 6.561(36)       & 6.281(32)        \\
 N302 & 36.219(49) & 36.295(48)  & 35.090(49)      & 35.667(49)       & 8.153(25)  & 7.305(22)   & 6.778(22)       & 6.496(20)        \\
 J303 & 36.496(37) & 36.579(31)  & 35.468(35)      & 36.039(29)       & 8.164(33)  & 7.401(29)   & 6.799(29)       & 6.586(26)        \\
 J304 & 36.485(28) & 36.568(28)  & 35.437(28)      & 36.015(27)       & --         & --          & --              & --               \\
 E300 & 36.603(18) & 36.697(16)  & 35.646(19)      & 36.221(16)       & 8.461(15)  & 7.613(13)   & 7.057(13)       & 6.780(12)        \\
 \hline
 J500 & 35.529(32) & 35.517(30)  & 35.008(29)      & 35.200(30)       & 7.118(59)  & 6.643(53)   & 6.499(56)       & 6.248(50)        \\
 J501 & 35.704(42) & 35.696(45)  & 35.227(40)      & 35.418(44)       & --         & --          & --              & --               \\
\hline
\end{tabular} 
\label{tab:aSDsub}
\end{table}

%% file: tables/tab_ls.tex
\begin{table}[!p]
\caption{Values of the $\Delta_{\rm ls}\aSDO$ contribution
in units of $10^{-10}$, for the local-local (${\scriptstyle\rm LL}$) and for the local-conserved (${\scriptstyle\rm CL}$) discretizations of the correlation function, as described in the main text.  
The finite-size correction has been applied.
}
\vskip 0.1in
\renewcommand{\arraystretch}{1.1}  
\begin{tabular}{
		l|@{\hskip 0.5em}c@{\hskip 0.5em}c@{\hskip 0.5em}|@{\hskip 0.5em}c@{\hskip 0.5em}c@{\hskip 0.5em}
	}
\hline
        & \multicolumn{2}{c|@{\hskip 0.5em}}{$-{\textstyle\frac{1}{3}}\Delta_{\rm ls}\aSDO$ - Set 1} 
        & \multicolumn{2}{c@{\hskip 0.5em}}{$-{\textstyle\frac{1}{3}}\Delta_{\rm ls}\aSDO$ - Set 2}
\\
\hline
id      &        ${\scriptstyle(LL)}$   &       ${\scriptstyle(CL)}$     &
                 ${\scriptstyle(LL)}$   &       ${\scriptstyle(CL)}$     
\\
\hline
  A654 & 0.048(14)  & 0.105(14)   & 0.435(15)       & 0.401(14)        \\
  \hline
  H102 & 0.044(10)  & 0.074(10)   & 0.257(10)       & 0.244(10)        \\
  H105 & 0.126(20)  & 0.187(20)   & 0.546(19)       & 0.520(19)        \\
  N101 & 0.147(05)  & 0.208(05)   & 0.562(03)       & 0.535(05)        \\
  C101 & 0.182(12)  & 0.264(11)   & 0.720(11)       & 0.691(10)        \\
  C102 & 0.223(08)  & 0.318(06)   & 0.823(06)       & 0.790(06)        \\
  D150 & 0.215(10)  & 0.315(07)   & 0.883(07)       & 0.848(06)        \\
  \hline
  S400 & 0.073(12)  & 0.097(12)   & 0.239(12)       & 0.234(12)        \\
  N451 & 0.172(04)  & 0.217(04)   & 0.468(04)       & 0.458(04)        \\
  D450 & 0.248(04)  & 0.309(03)   & 0.656(03)       & 0.643(03)        \\
  D451 & 0.299(05)  & 0.368(03)   & 0.754(03)       & 0.738(03)        \\
  D452 & 0.280(06)  & 0.350(05)   & 0.763(05)       & 0.746(05)        \\
  \hline
  N203 & 0.092(12)  & 0.110(12)   & 0.202(12)       & 0.204(11)        \\
  N200 & 0.212(13)  & 0.240(12)   & 0.409(12)       & 0.407(12)        \\
  D200 & 0.303(11)  & 0.343(11)   & 0.581(11)       & 0.578(10)        \\
  D201 & 0.356(06)  & 0.400(05)   & 0.660(05)       & 0.656(05)        \\
  E250 & 0.355(13)  & 0.402(12)   & 0.688(12)       & 0.685(11)        \\
  \hline
  N302 & 0.141(15)  & 0.148(14)   & 0.201(14)       & 0.200(14)        \\
  J303 & 0.301(12)  & 0.318(11)   & 0.416(12)       & 0.418(11)        \\
  J304 & 0.415(08)  & 0.435(08)   & 0.558(08)       & 0.558(07)        \\
  E300 & 0.419(05)  & 0.442(04)   & 0.573(05)       & 0.576(04)        \\
  \hline
  J501 & 0.163(12)  & 0.163(13)   & 0.196(12)       & 0.192(13)        \\
\hline
\end{tabular} 
\label{tab:aSDls}
\end{table}

%% file: tables/tab_lc.tex
\begin{table}[!p]
\caption{Values of the $\Delta_{\rm lc}\bSD$ contribution with $Q^2=25\,{\rm GeV}^2$
in units of $10^{-10}$, for the local-local (${\scriptstyle\rm LL}$) and for the local-conserved (${\scriptstyle\rm CL}$) discretizations of the correlation function, as described in the main text. The finite-size correction has been applied. }
\vskip 0.1in
\renewcommand{\arraystretch}{1.1}  
\begin{tabular}{
		l|@{\hskip 0.5em}c@{\hskip 0.5em}c@{\hskip 0.5em}|@{\hskip 0.5em}c@{\hskip 0.5em}c@{\hskip 0.5em}
	}
\hline
        & \multicolumn{2}{c|@{\hskip 0.5em}}{$-{\textstyle \frac{4}{9}} \Delta_{\rm lc}\bSD$ - Set 1} 
        & \multicolumn{2}{c@{\hskip 0.5em}}{$-{\textstyle \frac{4}{9}} \Delta_{\rm lc}\bSD$ - Set 2}
\\
\hline
id      &        ${\scriptstyle(LL)}$   &       ${\scriptstyle(CL)}$     &
                 ${\scriptstyle(LL)}$   &       ${\scriptstyle(CL)}$     
\\
\hline
 A653 & -0.194(28) & 1.936(16) & 4.886(14)     & 3.269(12)      \\
 A654 & -0.196(33) & 1.886(20) & 4.907(16)     & 3.305(14)      \\
 \hline
 H101 & 0.435(24)  & 2.147(15) & 3.854(15)     & 3.294(11)      \\
 H102 & 0.418(28)  & 2.122(17) & 3.887(17)     & 3.319(14)      \\
 N101 & 0.380(24)  & 2.074(14) & 3.886(14)     & 3.319(11)      \\
 C101 & 0.374(24)  & 2.058(15) & 3.912(15)     & 3.338(11)      \\
 \hline
 B450 & 0.857(23)  & 2.302(14) & 3.272(18)     & 3.275(12)      \\
 S400 & 0.963(24)  & 2.319(15) & 3.314(18)     & 3.314(13)      \\
 D450 & 0.806(20)  & 2.199(12) & 3.277(17)     & 3.270(11)      \\
 D452 & 0.768(20)  & 2.171(12) & 3.284(17)     & 3.270(11)      \\
 \hline
 N202 & 1.342(21)  & 2.446(14) & 2.790(17)     & 3.182(13)      \\
 N203 & 1.304(16)  & 2.418(11) & 2.772(15)     & 3.175(10)      \\
 N200 & 1.261(14)  & 2.367(09) & 2.756(13)     & 3.151(09)      \\
 D200 & 1.225(18)  & 2.329(14) & 2.742(16)     & 3.136(12)      \\
 E250 & 1.218(14)  & 2.309(09) & 2.746(14)     & 3.130(09)      \\
 \hline
 N300 & 1.823(17)  & 2.581(13) & 2.499(15)     & 3.018(12)      \\
 N302 & 1.761(11)  & 2.517(09) & 2.447(10)     & 2.968(09)      \\
 J303 & 1.832(16)  & 2.525(14) & 2.512(14)     & 2.987(13)      \\
 E300 & 1.692(11)  & 2.440(08) & 2.409(11)     & 2.919(07)      \\
 \hline
 J500 & 2.144(15)  & 2.647(11) & 2.472(15)     & 2.892(11)      \\
\hline
\end{tabular} 
\label{tab:aSDlc}
\end{table}

%% file: tables/tab_disc.tex
\begin{table}[!pt]
\caption{Values of the charm-disconnected contributions
in units of $10^{-10}$, for the conserved-conserved (${\scriptstyle\rm CC}$) discretization of the correlation function, as described in the main text. }
\vskip 0.1in
\renewcommand{\arraystretch}{1.1}  
\begin{tabular}{
		l|@{\hskip 0.5em}c@{\hskip 0.5em}c@{\hskip 0.5em}|@{\hskip 0.5em}c@{\hskip 0.5em}c@{\hskip 0.5em}
	}
\hline
        & \multicolumn{2}{c|@{\hskip 0.5em}}{${\textstyle \frac{4}{9}}\aSDf{\rm c}{\rm c}_{\rm disc}$} 
        & \multicolumn{2}{c@{\hskip 0.5em}}{${\textstyle \frac{2}{3\sqrt{3}}}\aSDf{\rm c}{\rm 8}_{\rm disc}$}
\\
\hline
id      &        Set 1   &       Set 2     &
                 Set 1   &       Set 2     
\\
\hline
  A654 & 0.0122(04)   & 0.0055(02)        & -0.0086(07)  & -0.0035(04)       \\
  \hline
  H102 & 0.0097(06)   & 0.0050(03)        & -0.0036(06)  & -0.0018(03)       \\
  H105 & 0.0112(06)   & 0.0058(03)        & -0.0114(12)  & -0.0059(08)       \\
  N101 & 0.0113(06)   & 0.0059(03)        & -0.0118(10)  & -0.0058(06)       \\
  C101 & 0.0104(08)   & 0.0054(04)        & -0.0151(19)  & -0.0072(11)       \\
  C102 & 0.0127(08)   & 0.0064(05)        & -0.0166(21)  & -0.0082(13)       \\
  D150 & 0.0119(12)   & 0.0063(07)        & -0.0211(49)  & -0.0119(30)       \\
  \hline
  S400 & 0.0101(05)   & 0.0058(03)        & -0.0051(05)  & -0.0030(04)       \\
  N451 & 0.0102(08)   & 0.0060(05)        & -0.0054(11)  & -0.0028(07)       \\
  D450 & 0.0109(11)   & 0.0063(07)        & -0.0122(26)  & -0.0070(17)       \\
  D451 & 0.0096(10)   & 0.0053(06)        & -0.0122(21)  & -0.0072(15)       \\
  D452 & 0.0091(08)   & 0.0050(05)        & -0.0114(27)  & -0.0060(18)       \\
  \hline
  N203 & 0.0078(10)   & 0.0052(07)        & -0.0043(07)  & -0.0029(06)       \\
  N200 & 0.0091(10)   & 0.0061(07)        & -0.0064(13)  & -0.0046(10)       \\
  D200 & 0.0070(10)   & 0.0045(07)        & -0.0074(20)  & -0.0049(16)       \\
  D201 & 0.0076(14)   & 0.0051(10)        & -0.0074(27)  & -0.0053(22)       \\
  E250 & 0.0092(09)   & 0.0062(06)        & -0.0095(24)  & -0.0057(20)       \\
  \hline
  N302 & 0.0064(19)   & 0.0049(16)        & -0.0004(15)  & -0.0001(14)       \\
  J303 & 0.0048(14)   & 0.0036(12)        & -0.0044(18)  & -0.0035(17)       \\
  J304 & 0.0056(12)   & 0.0043(10)        & -0.0041(16)  & -0.0028(15)       \\
  E300 & 0.0040(19)   & 0.0029(16)        & -0.0042(29)  & -0.0029(26)       \\
  \hline
  J501 & 0.0009(22)   & 0.0008(20)        & -0.0009(18)  & -0.0007(18)       \\
\hline
\end{tabular} 
\label{tab:aSDdisc}
\end{table}